\documentclass[namedreferences]{SolarPhysics}
\usepackage[optionalrh]{spr-sola-addons} 
\usepackage{graphicx}
\usepackage{color}
\usepackage{url}             

\newcommand{\aap}     {{\it Astron. Astrophys.}}

\newcommand{\apj}     {{\it Astrophys. J.}}
\newcommand{\apjl}    {{\it Astrophys. J. Lett.}}

\newcommand{\jgr}     {{\it J. Geophys. Res.}}

\newcommand{\nat}     {{\it Nature}}

\newcommand{\pasj}    {{\it Publ. Astron. Soc. Japan}}

\newcommand{\solphys} {{\it Solar Phys.}}

\newcommand{\ssr}     {{\it Space Sci. Rev.}}

\begin{document}
\begin{article}
\begin{opening}

\title{A Tiny Eruptive Filament as a Flux-Rope Progenitor and Driver of a Large-Scale CME and Wave}

\author{V.V.~\surname{Grechnev}$^{1}$\sep
        A.M.~\surname{Uralov}$^{1}$\sep
        A.A.~\surname{Kochanov}$^{1}$\sep
        I.V.~\surname{Kuzmenko}$^{2}$\sep
        D.V.~\surname{Prosovetsky}$^{1}\sep$
        Ya.I.~\surname{Egorov}$^{1}$\sep
        V.G.~\surname{Fainshtein}$^{1}$\sep
        L.K.~\surname{Kashapova}$^{1}$}

 \runningauthor{Grechnev et al.}
 \runningtitle{Eruptive filament as a driver of a CME}

\institute{$^{1}$ Institute of Solar-Terrestrial Physics SB RAS,
                  Lermontov St.\ 126A, Irkutsk 664033, Russia
                  email: \url{grechnev@iszf.irk.ru} email: \url{uralov@iszf.irk.ru}\\
           $^{2}$ Ussuriysk Astrophysical Observatory, Solnechnaya
                  St. 21, Primorsky Krai, Gornotaezhnoe 692533, Russia
                  email: \url{kuzmenko_irina@mail.ru}}

\date{Received ; accepted }

\begin{abstract}

A solar eruptive event SOL2010-06-13 observed with the
\textit{Atmospheric Imaging Assembly} (AIA) of the \textit{Solar
Dynamics Observatory} (SDO) has been extensively discussed in the
contexts of the CME development and an associated
extreme-ultraviolet (EUV) wave-like transient in terms of a shock
driven by the apparent CME rim. Continuing the analysis of this
event, we have revealed an erupting flux rope, studied its
properties, and detected wave signatures inside the developing
CME. These findings have allowed us to establish new features in
the genesis of the CME and associated EUV wave and to reconcile
all of the episodes into a single causally-related sequence. (1)~A
hot 11~MK flux rope developed from the structures initially
associated with a compact filament system. The flux rope expanded
with an acceleration of up to 3~km~s$^{-2}$ one minute before a
hard X-ray burst and earlier than any other structures, reached a
velocity of 420~km~s$^{-1}$, and then decelerated to about
50~km~s$^{-1}$. (2)~The CME development was driven by the
expanding flux rope. Closed coronal structures above the rope got
sequentially involved in the expansion from below upwards, came
closer together, and apparently disappeared to reveal their common
envelope, the visible rim, which became the outer boundary of the
cavity. The rim was probably associated with the separatrix
surface of a magnetic domain, which contained the pre-eruptive
filament. (3)~The rim formation was associated with a successive
compression of the upper active-region structures into the CME
frontal structure (FS). When the rim was formed, it resembled a
piston. (4)~The disturbance responsible for the consecutive CME
formation episodes was excited by the flux rope inside the rim,
and then propagated outward. EUV structures arranged at different
heights started to accelerate, when their trajectories in the
distance--time diagram were crossed by that of the fast front of
this disturbance. (5)~Outside the rim and FS, the disturbance
propagated like a blast wave, manifesting in a type II radio burst
and a leading part of the EUV transient. Its main, trailing part
was the FS, which consisted of swept-up 2~MK coronal loops
enveloping the expanding rim. The wave decelerated and decayed
into a weak disturbance soon afterwards, being not driven by the
trailing piston, which slowed down.

\end{abstract}
\keywords{Filament Eruptions; Coronal Mass Ejections; Shock Waves;
Type II Bursts}

\end{opening}

\section{Introduction}
  \label{S-Introduction}

A solar eruptive event in active region (AR) 11079 at S21~W82 was
observed on 13 June 2010 from about 05:30 to 05:50 (all times are
referred to UT) with the \textit{Atmospheric Imaging Assembly}
(AIA) on board the \textit{Solar Dynamics Observatory} (SDO;
\opencite{Lemen2012AIA}). Observations with the \textit{Sun Earth
Connection Coronal and Heliospheric Investigation} instrument
suite (SECCHI; \opencite{Howard2008}) on the
\textit{Solar-Terrestrial Relations Observatory} (STEREO;
\opencite{Kaiser2008}) from a different vantage point complement
the picture of the event. This event has been extensively
discussed in the contexts of the CME development
\cite{Patsourakos2010} and an associated extreme-ultraviolet (EUV)
wave-like transient in terms of a piston shock driven by the
apparent rim of the CME bubble (\opencite{Ma2011};
\opencite{Kozarev2011}; \opencite{Gopalswamy2012};
\opencite{Eselevich2012}; \opencite{Kouloumvakos2014}).
Nevertheless, some important questions remain unanswered. It is
still unclear where the flux rope was located, how it evolved, and
which properties it had. It is uncertain how the CME was formed
and what were the progenitors of its structural components.

\inlinecite{Patsourakos2010} followed the lift-off of the CME
bubble in the SOL2010-06-13 event. The authors detected an
eruptive filament, whose rise caused the rise of surrounding loops
which eventually formed an EUV cavity. They concluded that the
bubble was formed from a set of pre-existing loops during the main
flare phase, while the rise and possible instability of the
filament was considered as the possible CME trigger. The upper
limit for the size of a hypothetical pre-existing flux rope was
estimated to be very small, about 20~Mm.

\inlinecite{Eselevich2013} measured the expansion of the rising
coronal loops starting from $\approx 30$~Mm and found their
sequential involvement in the motion from below upwards during the
CME formation. Having not noticed the eruptive filament, the
authors proposed that the source of the CME was a magnetic tube
emerging with a high speed from below the photosphere.

In spite of high-resolution multi-wavelength observations by
SDO/AIA and \textit{Extreme UltraViolet Imager} (EUVI) on STEREO-A
and many efforts applied by several researchers, the flux rope
escapes detection so far. The origin and regime of the CME-related
wave remain conjectural. Based on a sophisticated thermodynamic
magnetohydrodynamic (MHD) model, \inlinecite{Downs2012} simulated
a detailed evolution of the EUV wave in this event in realistic
coronal conditions. The authors concluded that its outer,
propagating component had properties of a fast-mode wave, but
their analysis could not ascertain the wave excitation scenario.
It is still unclear where and how the presumable shock wave
developed.

Genesis of the flux rope, CME formation, and shock wave excitation
scenario are common long-standing issues for many similar events.
Addressing these issues promises reconciliation of existing
concepts with observational challenges and progress in
understanding eruptive events and underlying processes.

The basic guidelines to solve these problems are provided by the
standard flare model (`CSHKP'; \opencite{Car64};
\opencite{Sturrock66}; \opencite{Hirayama1974}; \opencite{Kopp76})
and its later developments. According to
\inlinecite{Hirayama1974}, the flare current sheet forms due to
the lift-off of a filament, whose eruption is driven by an MHD
instability of an increasing current in the filament. This can be
the torus instability governed by the Lorentz force
(\opencite{Anzer1978}; \citeauthor{Chen1989},
\citeyear{Chen1989,Chen1996}). A twisted flux rope can be formed
from an initial sheared configuration like a filament
\cite{vanBallegooijenMartens1989, Uralov1990,
InhesterBirnHesse1992, LongcopeBeveridge2007}. Observational
studies confirm the formation of flux ropes during flares (see,
\textit{e.g.}, \opencite{Asai2004}; \opencite{Qiu2007};
\opencite{Miklenic2009}) and their concurrent impulsive
acceleration (\opencite{Zhang2001}; \citeauthor{Temmer2008},
\citeyear{Temmer2008, Temmer2010}). The accelerating flux rope
must produce an MHD disturbance. Propagating into surrounding
regions, where the fast-mode speed is lower, the disturbance must
rapidly steepen into a shock (\citeauthor{Grechnev2011_I},
\citeyear{Grechnev2011_I, Grechnev2014_II};
\opencite{Afanasyev2013}), and then expand ahead of the CME like a
decelerating blast wave for some time. If the CME is slow, then
the shock eventually decays. Otherwise, the frontal part of the
shock changes to the bow-shock regime.

Different events, ranging from the GOES B-class up to the X-class,
exhibited these scenarios. The active role of filaments or similar
structures as progenitors of flux ropes (\opencite{Uralov2002};
\opencite{Grechnev2006erupt}) was confirmed by observations
(\opencite{ChenBastianGary2014}; \citeauthor{Grechnev2014_I},
\citeyear{Grechnev2014_I,Grechnev2015}). The excitation by such
structures of waves, which rapidly steepened into the shocks
manifesting in `EUV waves' and type II bursts, was first
demonstrated by \inlinecite{Grechnev2011_I} and then confirmed in
later studies of different events (see, \textit{e.g.},
\citeauthor{Grechnev2011_I}, \citeyear{Grechnev2013_20061213,
Grechnev2014_II, Grechnev2015}).

The 13 June 2010 event allows us to confront the outlined picture,
which we develop, with different views of the authors, who studied
this event previously, in order to verify our scenarios, specify
and elaborate the conjectures widely invoked. Our new in-depth
analysis of this eruptive event pursues the major unanswered
questions of the genesis of the flux rope and its properties; how
was the CME formed; where and how was the wave excited. We have
revealed the developing flux rope and the appearance of an
impulsively excited wave inside the forming CME, and studied some
of their properties.

The fact that the development of neither the flux rope nor the
shock wave have been detected previously indicates that various
temperature ranges should be examined using sensitive image
processing. Section~\ref{S-Methodical_Issues} addresses these
issues and our measurement techniques. Using them, we then analyze
the observations and discuss the results. Section~\ref{S-Overview}
considers the geometry of the CME bubble and the orientation of
the flux rope. Section~\ref{S-Flux_Rope} is devoted to the flux
rope. Section~\ref{S-CME_Formation} analyzes the CME formation.
Section~\ref{S-Wave} addresses the wave signatures.
Section~\ref{S-Discussion} discusses the origin of the observed
structures, compares the findings with a traditional view, and
presents an updated scenario of an eruptive event inferred from
the observations. Section~\ref{S-Conclusion} summarizes the
outcome from our analysis, outlines its implications, and finishes
with concluding remarks.

\section{Methodical Issues}
 \label{S-Methodical_Issues}

\subsection{SDO/AIA Images and their Processing}
 \label{S-image_processing}

The major observational data we use came from SDO/AIA. The
temperature response functions for the EUV channels of AIA were
presented by \inlinecite{Boerner2012}, \inlinecite{Lemen2012AIA},
and \inlinecite{Downs2012}. We use the well-known 171 and 193~\AA\
channels sensitive to normal coronal temperatures. The temperature
response of the 211~\AA\ channel resembles the major peak of the
193~\AA\ channel shifted to 2~MK but lacks a minor high-temperature
peak. The 94 and 131~\AA\ channels have two temperature sensitivity
windows. The lower-temperature windows are sensitive to normal
coronal temperatures, and the higher-temperature those have peaks at
6.3 and 10~MK, respectively. We also use the 304~\AA\ channel, which
is sensitive to temperatures around $5 \times 10^4$~K, with a lesser
contribution from hotter plasmas around 1.8~MK.

Due to the location of the active region close to the limb, the
erupting features were observed by AIA against the off-limb
background. It has a considerable diffuse component, whose
brightness is maximum at the limb and decreases with height. This
diffuse background substantially reduces the contrast of the
erupting features. In addition, their brightness, $I$,
dramatically decreases in their expansion; with a conserved number
of emitting particles, the brightness depends on a linear size,
$r$, as $I \propto r^{-5}$ \cite{Uralov2014, Grechnev2015}. Such
widely used ways as subtracting of an earlier image or dividing by
it do not reveal static features and contain traces of the base
image.

We have computed the radial brightness distributions from
averages over ten pre-event images in each AIA channel using a ring
scanning with an increasing radius (see \opencite{Kochanov2013}).
This way is still not perfect because of strong differences between
the low-latitude corona and regions above polar coronal holes.
Nevertheless, the azimuthally averaged distributions allowed us to
considerably enhance the appearance of off-limb features.

Subtraction of these background distributions enhances the
contrast of off-limb features. Dividing by these distributions
compensates for the upwards brightness decrease (probably, a
similar way was used by \inlinecite{Ma2011}). We use various
combinations of both these ways.

\subsection{Kinematic Measurements}
 \label{S-kinematic_measurements}

Properties of eruptive structures and the causal relations between
the underlying processes can be recognized from their kinematics,
which is basically inferred from distance--time measurements. An
obvious straightforward way to find the velocity and acceleration
is the differentiation of the experimentally measured
distance--time points. However, the measurements of an eruptive
feature, which is usually faint relative to associated flare
emission, are complicated by a rapid decrease of its brightness or
opacity that leads to considerable uncertainties. The irregular
appearance of the measured feature in the images causes a scatter
of the inferred velocities and accelerations. Even the modern
elaborations of the measurement techniques based on the direct
differentiation of the experimental distance--time points
(\textit{e.g.}, \opencite{Vrsnak2007}; \opencite{Temmer2010}) do
not overcome this difficulty completely, because the difference
between the measured and actual position is always unknown.

An alternative approach is based on the fitting an analytic
function to the measurements. Its major advantage is that the
kinematical plots are calculated by means of the integration or
differentiation of the analytic fit, thus providing a smooth
outcome, rather than the differentiation of the measurements, that
gives an intrinsically scattered result. If the kinematics of an
analyzed feature is basically understood and described
theoretically, then the problem is to compute the parameters of
the corresponding analytic function.

\inlinecite{Warmuth2001} proposed that fast Moreton waves observed
in the H$\alpha$ line and considerably slower ``EIT waves''
observed in EUV at larger distances were due to the same
decelerating fast-mode disturbances. To fit their propagation, the
authors attempted to use a 2nd-order polynomial and a power-law
fit. \inlinecite{Grechnev2008shocks} pointed out that a freely
propagating blast-wave-like shock, which spent its energy to sweep
up the plasma with a radial power-law density falloff, $n(x)
\propto x^{-\delta}$ ($x$ is the distance from the eruption
center), and extrude it from the volume it occupied previously,
indeed had a power-law kinematics, $x(t) \propto t^{2/(5-\delta)}$
\textit{vs.} time $t$. Practically, we estimate the wave onset
time, $t_0$ (\textit{e.g.}, from the analysis of a type II burst
or from distance--time measurements), and, referring to the
distance from the eruption center to one of the measured wave
fronts, $x_1$, at time $t_1$, adjust $\delta$ to reach a best fit
of the wave trajectory
\begin{eqnarray}
x(t) = x_1[(t-t_0)/(t-t_1)]^{2/(5-\delta)}.
 \label{E-pl_fit}
\end{eqnarray}
This simple approximation satisfactorily fits various wave
signatures such as ``EUV waves'', type II bursts, and leading
edges of fast CMEs (see, \textit{e.g.},
\citeauthor{Grechnev2011_I} \citeyear{Grechnev2011_I,
Grechnev2011_III, Grechnev2013_20061213}). The power-law density
model, $n(h) = n_0(h/h_0)^{-\delta}$, with $h$ being the height
from the photosphere, $n_0 = 4.1 \times 10^8$~cm$^{-3}$, $h_0 =
100$~Mm, and $\delta = 2.6$ is close to the equatorial Saito model
\cite{Saito1970} at $h \geq 260$~Mm within $\pm 30\%$, providing a
steeper density falloff at lesser heights. For the usage of the
power-law fit in the analyses of imaging data and dynamic radio
spectra see \inlinecite{Grechnev2014_II}.

For the kinematics of CME structures, which have been completely
formed and acquired maximum accelerations, we use analytic
equations obtained in a self-similar approach \cite{Uralov2005,
Grechnev2008shocks}. This approximation is based on the fact that
the relation between the propelling and retarding forces (magnetic
pressure and tension, plasma pressure, and gravity) applied to any
element of the expanding CME, established after the impulsive
acceleration stage, decreases by the same factor with an increase
of the distance from the expansion center (\textit{e.g.},
\opencite{Low1982}). This approximation applies, as long as the
aerodynamic drag of the solar wind has a minor importance,
\textit{i.e.}, until the regime of the plasma extrusion by the CME
bubble changes to the regime of the plasma flow around its outer
surface. The self-similar equations are complex and resemble
hyperbolic functions in behavior \cite{Grechnev2014_II}.

The kinematics of eruptive features during the impulsive
acceleration stage has not yet been well understood. In this case,
a more or less suitable analytic function can be chosen from
considerations based on the properties, which have already been
established. One knows \textit{a priori} that the initial velocity
is typically small or zero, the final velocity is nearly constant,
and the acceleration occurs impulsively within a certain time.
Considerable short-time variations of the acceleration and
velocity are not expected well after the impulsive acceleration
stage. The particular shape of the acceleration pulse does not
substantially affect the distance--time plot because of the double
integration. A bell-shaped acceleration pulse meets these
speculations. If a real distance--time plot considerably deviates
from the kinematics described with a single acceleration pulse,
then a combination of two (or more) pulses can be used.

This approach has been successfully used in several studies by
different authors (\textit{e.g.}, \opencite{Gallagher2003};
\opencite{Sheeley2007}; \opencite{WangZhangShen2009}) as well as
in our studies (\citeauthor{Grechnev2011_I}
\citeyear{Grechnev2011_I, Grechnev2013_20061213, Grechnev2014_I};
\opencite{Alissandrakis2013}). In our technique, the results of
the fit are used as a starting estimate of the parameters of the
acceleration, and then they are optimized to outline the eruption
in a best way. Our ultimate criterion is to follow the analyzed
feature as closely as possible in all of the images. The major
source of the errors is the uncertainty in following the same
moving feature, whose visibility progressively decreases.

The most important issue in our present study is the kinematics of
the flux rope, which was even difficult to detect. We fit the
measured projected heights of the flux rope, $h(t)$, with a smooth
function that accounts for its acceleration and deceleration
phases. We modified equation (1) from \inlinecite{Sheeley2007} to
the following two-pulse form:
\begin{eqnarray}
  h(t) = h(t_1) +
\frac{1}{2}(v_\mathrm{f}^{+}+v_{0}^{+})(t-t_1) +
\frac{1}{2}(v_\mathrm{f}^{+}-v_{0}^{+})\tau_1\ln\Big[\cosh(\frac{t-t_1}{\tau_1})\Big]+
\\ \nonumber
h(t_2) + \frac{1}{2}(v_\mathrm{f}^{-}+v_{0}^{-})(t-t_2) +
\frac{1}{2}(v_\mathrm{f}^{-}-v_{0}^{-})\tau_2\ln\Big[\cosh(\frac{t-t_2}{\tau_2})\Big].
\end{eqnarray}
\noindent Here $v_0$ and $v_\mathrm{f}$ are the initial and final
asymptotic values of velocity for the acceleration (+) and
deceleration ($-$) pulses, $t_1$ and $t_2$ are the acceleration
and deceleration center times, $h(t_1)$ and $h(t_2)$ are the
corresponding heights, and $\tau_1$ and $\tau_2$ are the
timescales of the acceleration and deceleration. The corresponding
velocity, $v(t)$, is
\begin{eqnarray}
  v(t) = \frac{1}{2}(v_\mathrm{f}^+ + v_0^+) +\frac{1}{2}(v_\mathrm{f}^+-
v_0^+)\tanh\Big(\frac{t-t_1}{\tau_1}\Big)+
\\ \nonumber
\frac{1}{2}(v_\mathrm{f}^- + v_0^-)
+\frac{1}{2}(v_\mathrm{f}^--v_0^-)\tanh\Big(\frac{t-t_2}{\tau_2}\Big).
\end{eqnarray}

The acceleration profile, $a(t)$, with the contributions from the
two pulses is
\begin{equation}
a(t) = \frac{v_\mathrm{f}^+ - v_0^+}{2\tau_1}\Big[ 1-\tanh^2 \Big(
\frac{t-t_1}{\tau_1} \Big) \Big]+
\\ \nonumber
\frac{v_\mathrm{f}^- - v_0^-}{2\tau_2}\Big[ 1-\tanh^2 \Big(
\frac{t-t_2}{\tau_2} \Big) \Big].
\end{equation}

We fit the analytic function $h(t)$ to the measured data using the
Levenberg--Marquardt least-squares minimization
\cite{Levenberg1944, Marquardt1963} implemented by C.B.~Markwardt
in the {\it SolarSoft MPFIT} package. To evaluate the confidence
intervals of the resulting fit, \textit{i.e.}, to estimate the
influence of the measurement errors on the inferred quantities
(the velocity and acceleration profiles), we use a technique
similar to the parametric bootstrap method. Numerous simulation
runs are carried out to produce a large number of data sets, in
which the measured data points are displaced by normally
distributed pseudo-random numbers with $\sigma$
corresponding to the measurement errors. Then we calculate the
variance for the parameters of the fit. In this way, we also
monitor the stability of the fit against noisy data.

The whole set of the parameters used in our fit is redundant,
while $v_\mathrm{f}^{+} = v_{0}^{-}$ appears to be sufficient. We
have to keep all of them to ensure a stable behavior of our
fitting software in its present implementation.

To study the evolution of an expanding feature, we adjust
the field of view of the images to keep its visible size fixed
according to the measured kinematics
(\citeauthor{Grechnev2014_II}, \citeyear{Grechnev2014_II,
Grechnev2015}). Residual trends can be detected in looking at a
movie composed from the resized images to improve the parameters
of the fit at the next iteration.

\subsection{Estimations of Plasma Parameters}
 \label{S-DEM_inversion}

Important information about eruptive structures and their
properties can be obtained from their temperature and emission
measure (EM). Qualitative judgments about the temperatures of
coronal structures observed by SDO/AIA can be inferred from their
appearance in different channels based on their temperature
response functions \cite{Boerner2012, Lemen2012AIA, Downs2012}. If
EM of an emitting structure is known, then, with a size found from
the images, its density and mass can be estimated.

The most general way is the inversion of the differential EM (DEM)
of the structures observed nearly simultaneously in different AIA
channels. Actually, the image in each next AIA channel is produced
12~s after the preceding one. Eruptive features of our interest
can acquire high speeds that causes their appreciable
displacements in the AIA images even during the relatively short
time intervals between them. To reduce the errors due to this
effect, we resize the images produced in all of the AIA channels
according to the measured kinematics of a feature in question,
as described in the preceding section. Practically, we
take in this case two sets of the AIA images, one
observed one time step before the time of interest, and the second
-- the next set, resize each image according to its observation
time, and then interpolate each pair of the images in each channel
to the required time. This procedure improves the co-registration
of the images.

In our analysis we use for reliability two different ways. These
are the regularized inversion technique and software developed by
\citeauthor{HannahKontar2012} (\citeyear{HannahKontar2012};
hereafter HK for brevity) and a faster algorithm developed by
\citeauthor{Plowman2013} (\citeyear{Plowman2013}; hereafter PKM).
Then we compare the results produced with the HK and PKM
algorithms.

\section{Geometry and Orientation}
 \label{S-Overview}

\begin{figure} 
  \centerline{\includegraphics[width=\textwidth]
   {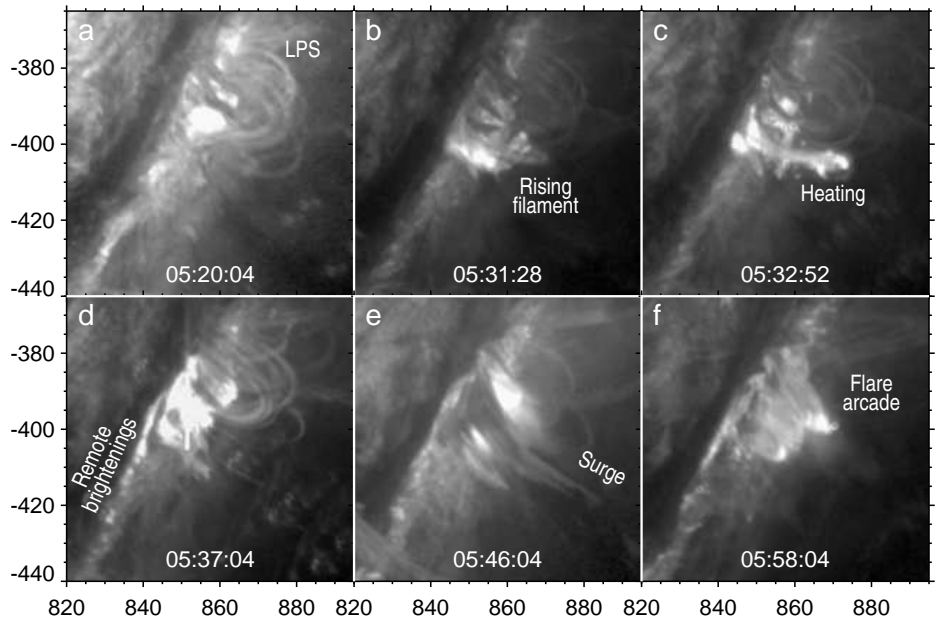}
  }
  \caption{Activity in the low corona throughout the event observed in SDO/AIA
304~\AA\ images. Label `LPS' in panel (a) denotes a loop prominence
system. The axes show hereafter the coordinates in arcsec from the
solar disk center.}
  \label{F-aia304}
  \end{figure}

Figure~\ref{F-aia304} presents some episodes of the event observed
in the SDO/AIA 304~\AA\ channel. In a pre-event configuration in
Figure~\ref{F-aia304}a, a loop prominence system (LPS) is
considerably inclined to the line of sight. LPSs are known to be
located above the main magnetic polarity inversion (neutral)
lines.

A rising filament in Figures \ref{F-aia304}b and \ref{F-aia304}c
becomes bright, which indicates its heating. Then the hot top of
the filament becomes transparent and disappears later on. In
Figure~\ref{F-aia304}d, remote compact bright kernels
intermittently appear and fade, being arranged in a direction
nearly parallel to the axis of the LPS. The plane of a dark surge
in Figure~\ref{F-aia304}e and the orientation of the flare arcade
in Figure~\ref{F-aia304}f also correspond to the inclined
direction of the LPS.

\subsection{Overall Configuration}

\begin{figure} 
  \centerline{\includegraphics[width=\textwidth]
   {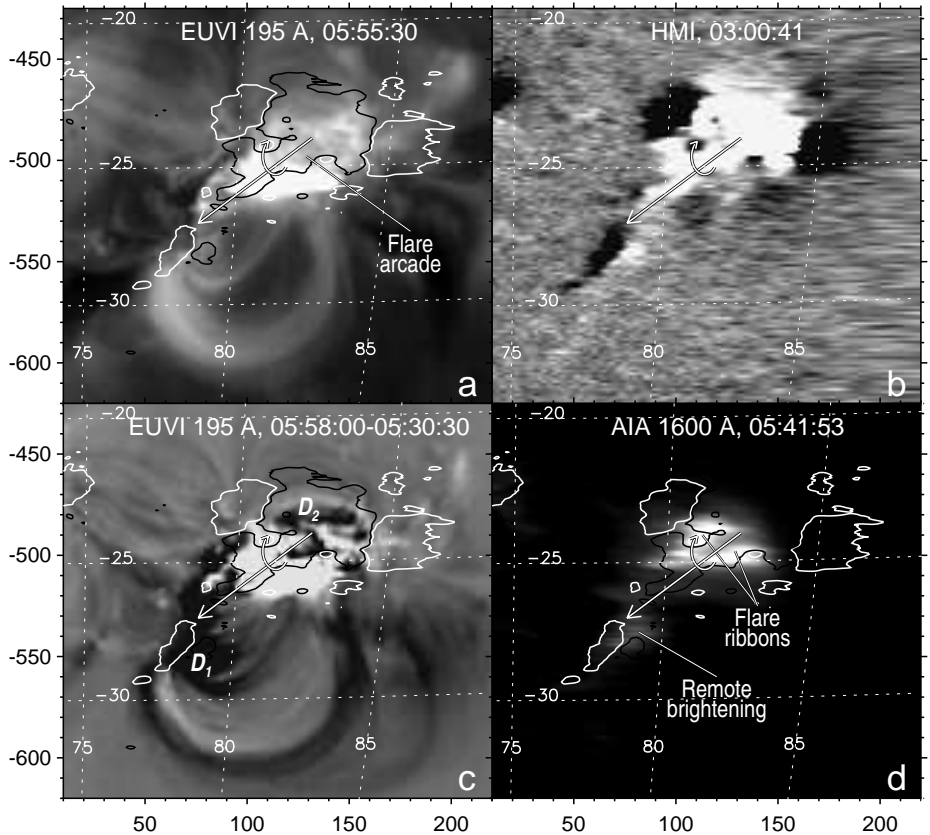}
  }
  \caption{Flare configuration observed from STEREO-A
along with contours of an SDO/HMI magnetogram transformed to this
viewing direction. The contour levels correspond to $-35$~G
(white) and $+35$~G (black) in the magnetogram smoothed with a
5-pixel boxcar. (a)~Flare arcade in an EUVI 195~\AA\ image.
(b)~HMI magnetogram, rotated to match the view from STEREO-A,
within a range of $\pm 50$~G. (c)~Dimmed regions in an EUVI
195~\AA\ difference image. The dimming regions presumably
associated with the footprints of the flux rope are denoted $D_1$
and $D_2$. (d)~Flare ribbons and remote brightening in an SDO/AIA
1600~\AA\ image transformed to the viewing direction from
STEREO-A. The straight arrow indicates the orientation of the
axial magnetic field in the flux rope ($-37^{\circ}$). The round
arrow shows the direction of the magnetic field in the arcade. The
heliographic grid corresponds to viewing from Earth.}
  \label{F-euvi_hmi_aia1600}
  \end{figure}

Complementary observations from the STEREO-A vantage point make
the overall configuration clearer. Figure~\ref{F-euvi_hmi_aia1600}
compares the STEREO-A/EUVI 195~\AA\ images with an SDO/HMI
line-of-sight magnetogram (Figure~\ref{F-euvi_hmi_aia1600}b) and
the flare ribbons in the AIA 1600~\AA\ image
(Figure~\ref{F-euvi_hmi_aia1600}d). The SDO data were transformed
to the viewing direction from STEREO-A. A strong projection
shrinkage of the near-the-limb flare site reduces the quality of
the transformed SDO images. The westernmost part of the
magnetogram appears with an inverted polarity, which is most
likely an artifact due to projection effects on magnetic field
inclined to the line of sight, which is a common feature in
magnetograms close to the limb.

The orientation of the flare arcade (partly saturated) in a late
EUVI image in Figure~\ref{F-euvi_hmi_aia1600}a corresponds to the
flare ribbons in a transformed 1600~\AA\ AIA image in
Figure~\ref{F-euvi_hmi_aia1600}d. The ribbons must be separated by
the magnetic neutral line, whose direction is shown by the
straight tilted arrow, which corresponds to the orientation of the
LPS and arcade in Figure~\ref{F-aia304}.

A difference image in Figure~\ref{F-euvi_hmi_aia1600}c reveals the
regions of dimming, some of which are probably due to
displacements of the loops visible in
Figure~\ref{F-euvi_hmi_aia1600}a or eruption of their neighbors.
The core dimmings D$_1$ and D$_2$ might be associated with the
footprints of an erupted flux rope \cite{HudsonWebb1997,
SterlingHudson1997, Webb2000, Mandrini2005}. This assumption is
confirmed by a remote brightening in 1600~\AA\ within D$_1$ in
Figure~\ref{F-euvi_hmi_aia1600}d (\textit{cf.}
Figure~\ref{F-aia304}d). A conjugate footpoint of the flux rope
must be within an opposite polarity; region D$_2$ meets this
requirement. The direction of the flux rope's azimuthal magnetic
field (the round arrow) should correspond to the flare arcade,
being prompted by the magnetogram in
Figure~\ref{F-euvi_hmi_aia1600}b, although considerably distorted.

The observations indicate that the flux rope was compact, with a
length comparable to its width. Its axis was initially tilted by
$\approx -37^{\circ}$ to the East.

\subsection{CME Lift-off}
 \label{S-CME_lift-off}

The eruption produced a CME, whose lift-off was observed by
SDO/AIA in different-temperature channels. They reveal various
coronal structures (see also the \url{AIA_131_171_loops.mpg} movie
in the electronic supplementary material).

\begin{figure} 
  \centerline{\includegraphics[width=\textwidth]
   {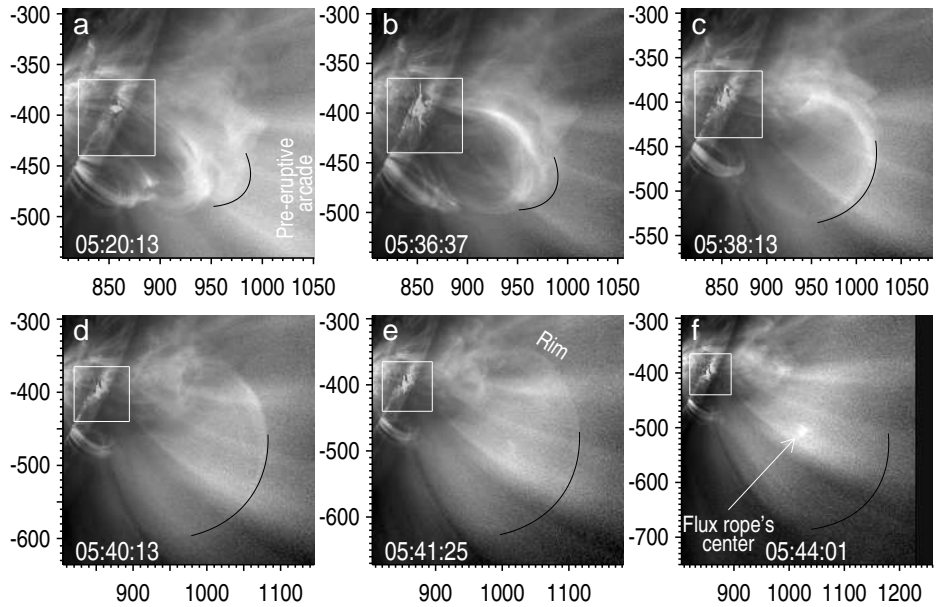}
  }
  \caption{Development of the CME bubble in SDO/AIA 171~\AA\ images.
The images are progressively resized to keep the size of the
outermost loop outlined by the black arc unchanged. The white
frame corresponds to the field of view in Figure~\ref{F-aia304}.}
  \label{F-bubble171}
  \end{figure}

The 171~\AA\ channel is sensitive to coronal features of
relatively low temperatures. Figure~\ref{F-bubble171} shows some
episodes of the CME development in the AIA 171~\AA\ images
starting from the pre-event configuration in
Figure~\ref{F-bubble171}a. The orientation of the pre-eruptive
arcade here corresponds to the inferred tilt of the flux rope's
axis of $-37^{\circ}$. The black arc outlines the top of the
visible set of the loops.

As the developing CME lifts off, the arcade loops get involved in
the expansion from below upwards \cite{Patsourakos2010,
Eselevich2013}. The rising lower loops press the overlying ones,
and finally all of them apparently merge into a very thin, nearly
circular rim. This process is addressed in
Section~\ref{S-CME_Formation}. Most authors, who studied this
event previously, invoked a traditional assumption of the identity
of the flux rope with the CME cavity, and related its outer
boundary to the visible rim. The rim was considered as a cross
section of the flux rope viewed nearly along its axis.

However, the rim in Figure~\ref{F-bubble171}e resembles a balloon
with a thin skin or a soap bubble. With the estimated orientation of
the flux rope, its cylindrical shape would contradict the appearance
of the rim. Even though a possible rotation of the flux rope around
the continuation of the solar radius is not excluded, the appearance
of the rim corresponds to a roughly spheroidal or pyriform
CME bubble rather than a long cylinder viewed nearly along its
axis.

The curvature of the black arc outlining the top of the arcade,
which transformed into the rim, decreased in Figures
\ref{F-bubble171}a--\ref{F-bubble171}d. This effect was revealed
by \inlinecite{Patsourakos2010}. Then the curvature gradually
increased again, as the movie demonstrates. A bright kernel inside
the bubble (visible also in 193 and 211~\AA) might be a
largest-opacity flux-rope center.

\begin{figure} 
  \centerline{\includegraphics[width=\textwidth]
   {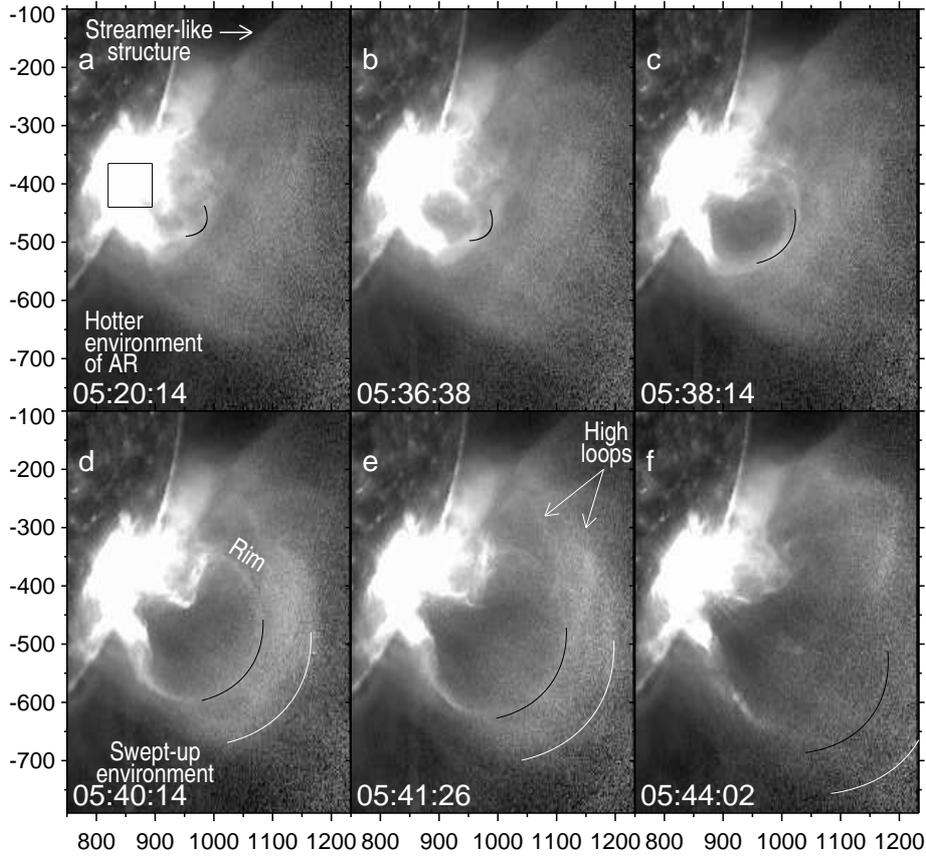}
  }
  \caption{Development of the CME bubble in SDO/AIA 211~\AA\ images.
The black arc corresponds to the arc in Figure~\ref{F-bubble171}.
The white arc in panels (d) -- (f) is $90^{\prime \prime}$ farther.
The expanding rim sweeps up the bright structures enclosed between
the arcs. The black frame in panel (a) corresponds to the field of
view in Figure~\ref{F-aia304}.}
  \label{F-bubble211}
  \end{figure}

The 211~\AA\ AIA images with a reduced background in
Figure~\ref{F-bubble211} reveal a low-brightness
higher-temperature environment of the AR consisting of closed
structures. They were higher and had different orientations from
the loops visible in 171~\AA. The loops in 211~\AA\ are faint in
the pre-event image in Figure~\ref{F-bubble211}a and become
discernible later (\textit{e.g.}, in Figure~\ref{F-bubble211}e).
The different orientations of the loops below the rim and above it
suggest its association with a separatrix surface.

A separatrix surface prevents mixing magnetic structures, which
belong to different magnetic domains isolated by this surface. The
expanding rim associated with a separatrix surface limits the
expansion of the loops below and sweeps up the structures above,
leaving a rarefied volume behind. No dimming is pronounced in
Figure~\ref{F-bubble211}, because we only reduced the
structureless coronal background in front of the bubble and behind
it, without subtracting any preceding image.

The black arc outlines the top of the rim, and the white arc,
whose radius is $90^{\prime \prime}$ larger, acceptably matches
the outer edge of the pileup, which probably becomes the
leading edge of the future CME later on. The expansion velocities
of the rim and pileup from 05:40 to 05:44 were not much different.

North of the AR, a ray-like feature resembling a small streamer is
denoted in Figure~\ref{F-bubble211}a. Presumably at its base, a
quadrupole configuration is present in STEREO-A/EUVI 195~\AA\ images
about $100^{\prime \prime}$ west of the AR. This site is a candidate
for a source region of a type~II radio burst discussed in
Section~\ref{S-Wave}.

\section{Flux Rope}
 \label{S-Flux_Rope}

\subsection{Genesis and Expansion}
  \label{S-Genesis_Expansion}

According to \inlinecite{Patsourakos2010}, the CME lift-off was
possibly triggered by the eruption of a tiny filament ($\lsim
20$~Mm). It is shown in Figures \ref{F-aia304}b and
\ref{F-aia304}c. We start to search for the elusive flux rope from
the activation of the filament observed in 131~\AA.

\begin{figure} 
  \centerline{
   \includegraphics[width=0.85\textwidth]{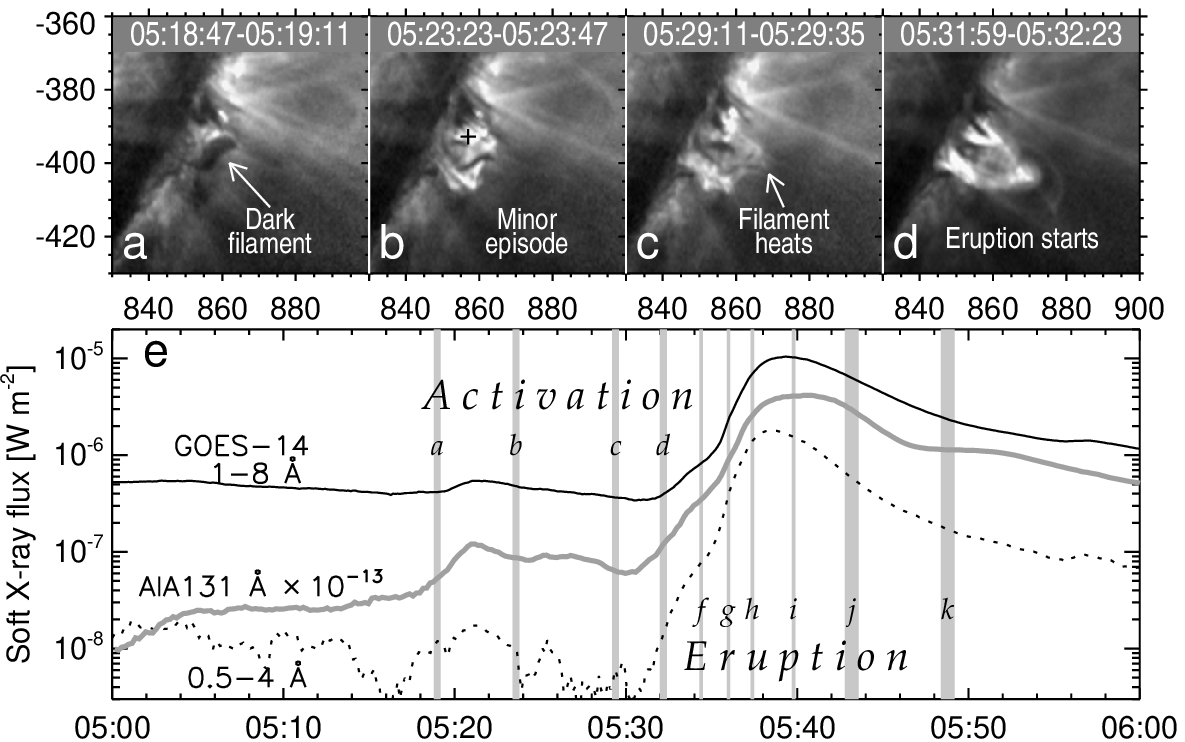}}
   \centerline{
      \includegraphics[width=0.85\textwidth]{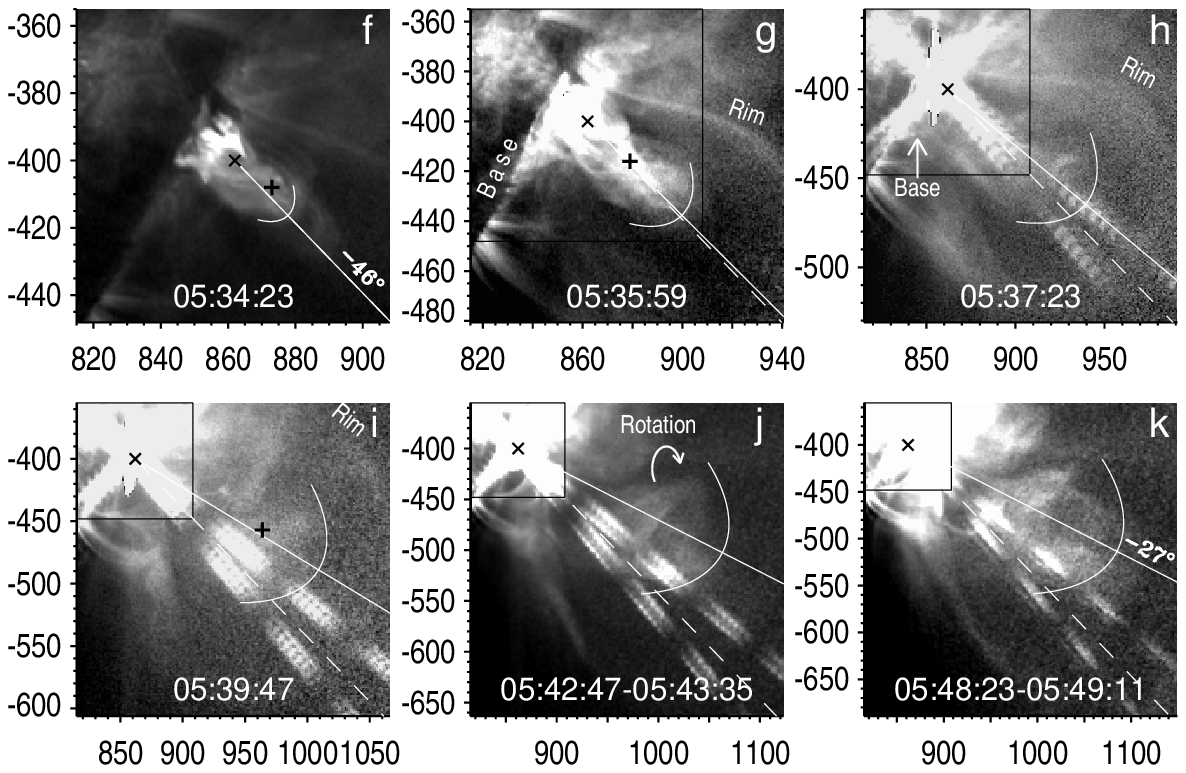}
  }
  \caption{(a--d)~Filament activation in averaged AIA 131~\AA\
images. (e)~GOES flux in 1--8~\AA\ (black solid) and 0.5--4~\AA\
(dotted) and a light curve computed from the 131~\AA\ images over
the flare region (gray). (f--k)~Erupting flux rope in 131~\AA\
images resized to keep its extent. The arcs outline its top. The
solid line goes from the origin of the measurements (slanted
cross) and the flux rope center. The dashed line shows its initial
orientation $-46^{\circ}$ from the west, the straight crosses in
panels (b), (f), (g), and (i) denote the positions, for which DEM
in Figure~\ref{F-rope_dem_profiles} was computed. The black frame
denotes the field of view in panel (f). The images in panels (j)
and (k) were averaged in the specified intervals after resizing.
The coordinates indicate position in arcsec from disk center at
the middle of the averaging intervals.}
  \label{F-flux_rope}
  \end{figure}

The initial dark filament (Figure~\ref{F-flux_rope}a) activated in
Figures \ref{F-flux_rope}b--\ref{F-flux_rope}d. The history of the
event can be followed in Figure~\ref{F-flux_rope}e from the two
soft X-ray (SXR) GOES channels and a light curve computed from the
131~\AA\ images over a $153^{\prime \prime} \times 153^{\prime
\prime}$ region centered at $(855^{\prime \prime}, -396^{\prime
\prime})$ to encompass the flare site. The gray vertical bars in
Figure~\ref{F-flux_rope}e represent the intervals, in which the
images were averaged. The 131~\AA\ light curve reveals a long
gradual rise of the emission during 05:00--05:30 from the filament
or its environment, indicating heating processes. This rise is not
present in the GOES data, which respond to the whole Sun's
emission.

The activated part of the filament in Figure~\ref{F-flux_rope}b
brightens up during a minor episode of 05:20--05:24 before the
major eruption, supporting its heating. A similar appearance of
the brightened filament in different AIA channels suggests a wide
range of plasma temperatures in its body. A response to this
episode in 1--8~\AA\ and even in 0.8--4~\AA\ (although marginal)
indicates that the filament brightening could be caught in the
high-temperature window of the 131~\AA\ channel. The top of the
filament in Figure~\ref{F-flux_rope}c becomes bright and
transparent. The eruption starts in Figure~\ref{F-flux_rope}d.

The average temperature of the brightened filament estimated from
the two GOES channels is $\approx 6.6$~MK. All of the estimates
indicate that the 131~\AA\ images are most promising to reveal the
disappearing top part of the erupting filament, its possible
relation to the flux rope, and the flux rope itself. The
high-temperature window of the 94~\AA\ channel could also be
appropriate, but its sensitivity is considerably lower. We
therefore focus on the 131~\AA\ images.

A sharp increase of the emission after 05:30
(Figure~\ref{F-flux_rope}e) caused strong overexposure effects
such as saturation, blooming, and oblique diffraction patterns.
Nevertheless, the image processing described in
Section~\ref{S-image_processing} allowed us to detect an erupting
flux rope in 131~\AA\ shown in Figures
\ref{F-flux_rope}f--\ref{F-flux_rope}k and the
\url{flux_rope_131.mpg} movie. The images are resized to fix the
visible extent of the flux rope using the kinematical measurements
described in the next section.

The heated filament body in Figure~\ref{F-flux_rope}d transforms
into an erupting bundle of twisted loops in
Figure~\ref{F-flux_rope}f. Faint outermost loops disappear soon.
The bundle rapidly expands along the dashed line inside the rim,
as a decreasing black frame (field of view in
Figure~\ref{F-flux_rope}f) indicates. Several threadlike loops are
rooted at the base denoted in Figures \ref{F-flux_rope}g and
\ref{F-flux_rope}h. As the rope rises, its Earth-facing base
expands southeast, producing the remote birghtenings in Figures
\ref{F-aia304}d and \ref{F-euvi_hmi_aia1600}d.

The direction of the lift-off (solid line) gradually turns aside by
$\approx 20^\circ$. The flux rope rotates (see the circular arrow in
Figure~\ref{F-flux_rope}j and the movie). More loops still erupt and
join the flux rope in latest images. The latest visible loops are
apparently injected into the northern part of the flux rope's
bottom.

This sequence of events is faintly visible in 131~\AA\
(characteristic temperature 10~MK). To reveal the flux rope in
Figure~\ref{F-flux_rope}, we had to average a few images within
the specified intervals. The flux rope can also be detected in
94~\AA\ (6.3~MK), but still poorer. These circumstances indicate
that its temperature was around 10~MK.

\subsection{Kinematics}

We used a few different ways to measure the kinematics of the
expanding flux rope. The major difficulty was its decreasing
brightness, which became comparable with noise in the images, so
that the flux rope eventually disappeared. To get a hint at its
final speed, we analyzed the images of the SOHO's \textit{Large
Angle and Spectroscopic Coronagraph} (LASCO;
\opencite{Brueckner1995}). The flux-rope's center expanded in
LASCO/C2 images after 09:30 with a speed of $\approx
52$~km~s$^{-1}$, while the speed of its top was considerably less
than the asymptotic speed of the frontal structure,
190~km~s$^{-1}$ (Figure~\ref{F-lasco} in Section~\ref{S-lasco}).
The top of the flux rope was 30--40\% ahead of its center visible
in late AIA 193~\AA\ images. These speculations indicate a final
speed of 60--80~km~s$^{-1}$ for the flux rope's top.

\begin{figure} 
  \centerline{\includegraphics[width=\textwidth]
   {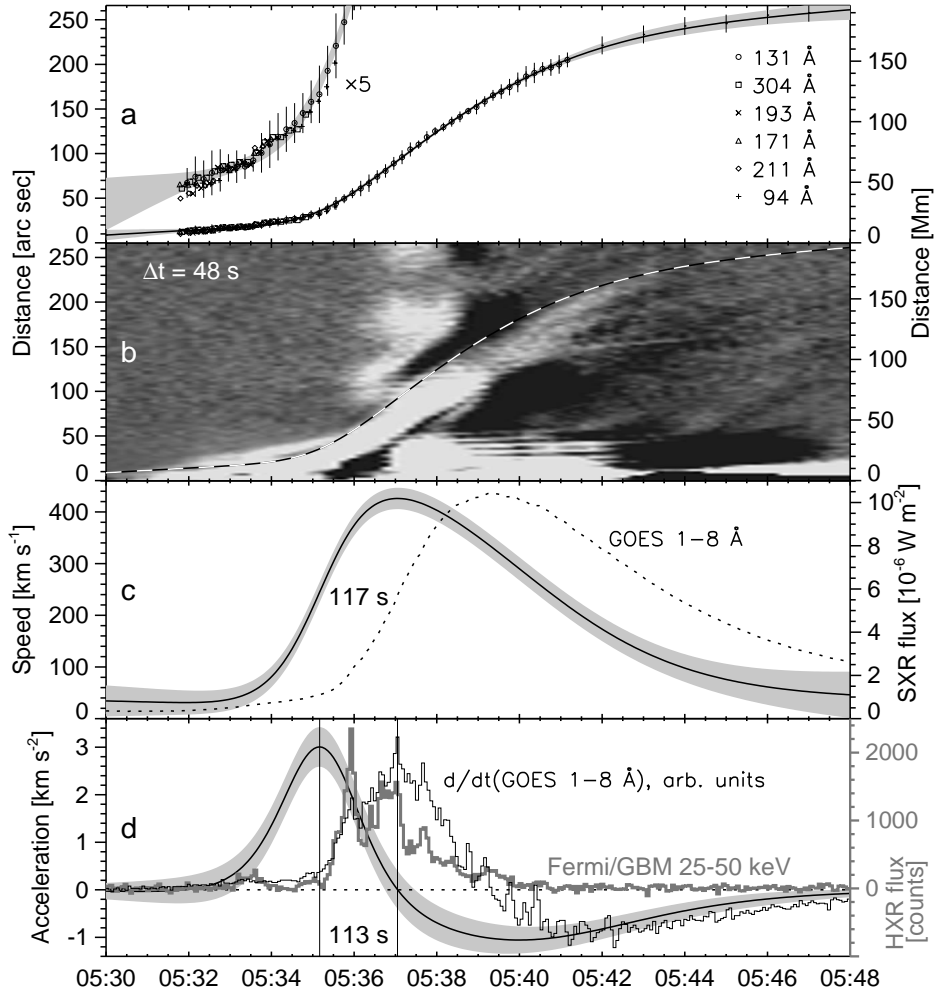}
  }
  \caption{Kinematics of the flux rope. (a)~Direct manual
distance--time measurements (symbols) and the analytic fit (solid
line). The initial part enlarged by a factor of five is also
shown. The bars represent the errors of measurements from the
131~\AA\ images estimated subjectively. The shadings in panels
(a), (c), and (d) represent the uncertainties evaluated by the
parametric fit. (b)~One-dimensional time history of the flux
rope's lift-off in 131~\AA\ running-difference images separated by
48~s. The dashed line represents the analytic fit from panel (a).
(c)~Velocity--time plot computed from the analytic fit (solid)
along with a GOES flux (dotted). (d)~Acceleration of the flux rope
(black), hard X-ray burst (gray), and the derivative of the GOES
flux (thin).}
  \label{F-rope_measurements}
  \end{figure}

Using the results of preliminary measurements, we produced a
movie, in which we scaled the field of view to compensate for the
expansion of the flux rope. The movie shows it clearer. If the
visible size of the flux rope in the movie still varied, then we
refined our measurements and repeated the attempt. These various
ways converged into the results presented in
Figure~\ref{F-rope_measurements}.

The starting manual distance--time measurements are presented in
Figure~\ref{F-rope_measurements}a. The heights are related to the
varying direction mentioned in the preceding section. The
correspondence between the tops of the filament measured from six
AIA channels while it was detectable (the magnified initial part)
and the flux rope in the 131~\AA\ images confirms their genetic
relation.

We also manually adjusted an acceleration profile consisting of a
positive pulse followed by a negative one \cite{Grechnev2015}.
Finally, elaborating this approach, we developed the automatic fit
described in Section~\ref{S-kinematic_measurements}. The black
curve in Figure~\ref{F-rope_measurements}a is its result. The
shading represents the calculated uncertainties.

The fit is superimposed on a time-history image in
Figure~\ref{F-rope_measurements}b (similar to the slit images used
by \opencite{Ma2011}; see also \opencite{Alissandrakis2013};
\opencite{Grechnev2014_I}). Each column of this image is a spatial
profile computed as averages over a 12-pixel ($7.2^{\prime
\prime}$) wide slice extracted from a running-difference 131~\AA\
image with a cadence of 48~s. The instant orientations of the
slices follow the turning flux rope. The expanding flux rope
appears in this image as a bright strip. The fit should be its
upper envelope. A bright feature visible above the fit from
05:31:30 till 05:36:00 is due to disappearing outermost loops seen
in Figure~\ref{F-flux_rope}f. They started expanding earlier and
had a nearly constant speed.

Figure~\ref{F-rope_measurements}c shows the velocity--time plot
computed from the analytic fit, the uncertainties (shaded), and
the GOES SXR flux. The flux-rope velocity started to sharply rise
at 05:33, exceeded 400~km~s$^{-1}$ at 05:37, and then decreased to
$\approx 50$~km~s$^{-1}$. The SXR flux is similar to the velocity
plot, but lags behind it by 117~s. The similarity of the rise
phases is expected in the scenario of \inlinecite{Hirayama1974},
where flare processes are driven by the erupting filament. The
similarity of the declining-phase parts might be due to expansion
of the flare arcade in the wake of the CME
\cite{LivshitsBadalyan2004}.

Figure~\ref{F-rope_measurements}d shows the computed acceleration
plot (thick black) with uncertainties, hard X-ray (HXR) flux from
the \textit{Gamma-Ray Burst Monitor} of the \textit{Fermi
Gamma-ray Space Telescope} (\textit{Fermi}/GBM;
\opencite{Meegan2009}) and reconstructed within the 25--50 keV
energy band for this burst (thick gray); and the derivative of the
GOES flux (thin). The flux rope underwent a strong acceleration up
to $3$~km~s$^{-2} \approx 11g_{\odot}$ at 05:35:10; $g_{\odot} =
274$~m~s$^{-2}$ is the solar gravity acceleration at the
photospheric level. The acceleration changed to a longer
deceleration, which reached $-1$~km~s$^{-2}$ at about 05:40:00.

The sharply accelerating flux rope must have produced a strong
wavelike disturbance. It must propagate omnidirectionally,
initially with a fast-mode speed, $V_\mathrm{fast}$. Typically,
$V_\mathrm{fast} \gsim 10^3$~km~s$^{-1}$ in the low corona above
active regions.

The time resolution of the \textit{Fermi}/GBM data of an enhanced
spectral resolution we use is 4~s. The time bins of the GOES-14
SXR data are 2~s. The positive portion of the derivative of the
SXR flux roughly resembles the HXR burst (the Neupert effect;
\opencite{Neupert1968}) and contains counterparts of most HXR
features without a detailed correspondence. The derivative of the
SXR flux is similar to the acceleration pulse, lagging behind by
113~s. The lag of the HXR and microwave emissions by 1--2 minutes
behind the acceleration of an eruptive structure seems to be a
systematic phenomenon. We observed it previously in a few other
events (\citeauthor{Grechnev2011_I} \citeyear{Grechnev2011_I,
Grechnev2013_20061213, Grechnev2015}).

\subsection{Differential Emission Measure}

\begin{figure} 
  \centerline{\includegraphics[width=0.8\textwidth]
   {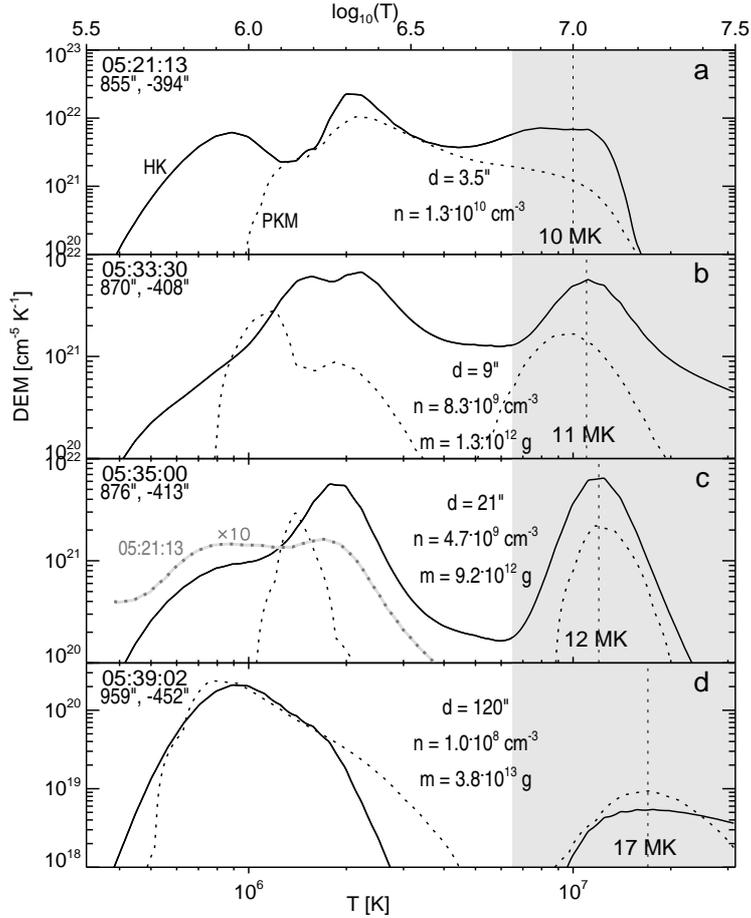}
  }
  \caption{DEM temperature profiles in the heated filament
during the minor filament heating episode at 05:21:13 (a) and the
flux rope at 05:33:30 (b), 05:35:30 (c), and 05:39:02 (d). DEM was
computed using the HK (solid) and PKM (dotted) methods. The plots
are related to the regions, whose centers are denoted by the
straight crosses in Figures \ref{F-flux_rope}b,
\ref{F-flux_rope}f, \ref{F-flux_rope}g, and \ref{F-flux_rope}i,
respectively. Their coordinates are listed in the panels. The gray
PKM profile in panel (c) magnified by a factor of 10 corresponds
to the same position before the eruption. The sensitivity of the
HK method is insufficient for this faint region. The densities and
masses were computed for the sizes, $d$, specified in the panels,
within the shaded temperature range.}
  \label{F-rope_dem_profiles}
  \end{figure}

We computed DEM from sets of AIA images produced nearly
simultaneously in different channels (see
Section~\ref{S-DEM_inversion}).
Figure~\ref{F-rope_dem_profiles} presents the results for
four episodes of the flux-rope development: the minor episode of the
filament heating in Figure~\ref{F-flux_rope}b (05:21:13,
Figure~\ref{F-rope_dem_profiles}a), the early flux-rope appearance
in Figure~\ref{F-flux_rope}f (05:33:30,
Figure~\ref{F-rope_dem_profiles}b), is half-height acceleration in
Figure~\ref{F-flux_rope}g (05:35:30,
Figure~\ref{F-rope_dem_profiles}c), and during the deceleration
stage in Figure~\ref{F-flux_rope}i (05:39:02,
Figure~\ref{F-rope_dem_profiles}d). The times and centers of the
boxes, in which DEM was calculated, are listed in the upper-left
corner of each panel. It was not possible to relate the calculations
to the same part of the flux rope because of sharp changes in its
shape and strong overexposure distortions of the AIA images. The
observation times, for which DEM was computed, are not identical to
the most representative images of the flux rope in
Figure~\ref{F-flux_rope}.

The PKM profiles in the lower-temperature range in Figures
\ref{F-rope_dem_profiles}b and \ref{F-rope_dem_profiles}c do not
look perfect, possibly due to the strong instrumental distortions,
while the HK profiles seem to be more plausible. Our major
interest is related to the shaded high-temperature domain. Here
both methods supplied similar results, differing quantitatively in
the maximum DEM by factors of 3.4, 2.9, and 0.58 for the three
times, respectively. The temperature of the flux rope
progressively increased from $\approx 10$~MK at 05:33:30 to 12~MK
at 05:35:00 and then to 17~MK at 05:39:02.

The visible widths of the flux rope at 05:35:00 and 05:39:02 were
$d_2 \approx 21^{\prime \prime}$ and $d_3 \approx 120^{\prime
\prime}$. If the total number of emitting particles inside the
expanding volume was conserved, then the expected decrease of the
brightness (\textit{i.e.}, DEM; \opencite{Grechnev2015}) should be
$(d_3/d_2)^5 \approx 6100$. Actually, the DEM decrease from
05:35:00 to 05:39:02 was much less, 1200 (HK method) and 230 (PKM
method). A similar situation occurred also between 05:33:30 and
05:35:00.

The density and mass of the hot flux rope listed in Figures
\ref{F-rope_dem_profiles}b--\ref{F-rope_dem_profiles}d were
estimated from its plane-of-the-sky width assuming the spherical
shape of its upper part (similar in
Figure~\ref{F-rope_dem_profiles}a). The ratios of their estimates
with the HK and PKM methods are 2.1, 1.1, and 0.89 at the three
times that seems to be reasonable, considering the faintness of
the flux rope.

All of the estimates indicate that the mass of the flux rope
considerable increased. Along with the increase of its
temperature, this fact suggests an ongoing injection of hot plasma
from the flaring region; otherwise, the temperature increase were
challenging. As mentioned, the injection of high-temperature loops
is indeed faintly visible at late stages of the eruption. The
rotational momentum supplied by these non-centrally injected loops
possibly caused the rotation of the flux rope indicated in
Figure~\ref{F-flux_rope}j and visible in the movie.

\section{Development of CME Structural Components and Their Later Expansion}
 \label{S-CME_Formation}

\subsection{CME Formation in EUV Images}
 \label{S-CME_formation_in_EUV}

The phenomena observed during the CME lift-off in this event were
analyzed by \inlinecite{Patsourakos2010};
\inlinecite{Eselevich2013}; and partly by \inlinecite{Ma2011}.
Having not detected the eruptive flux rope, the authors
nevertheless established a sequential involvement of coronal loops
in the motion from below upwards during the CME lift-off. Here we
study the relation between the erupting flux rope and the CME
formation.

\subsubsection{Rim and Inner Structures in 193~\AA\ AIA Images}

\begin{figure} 
  \centerline{\includegraphics[width=0.9\textwidth]
   {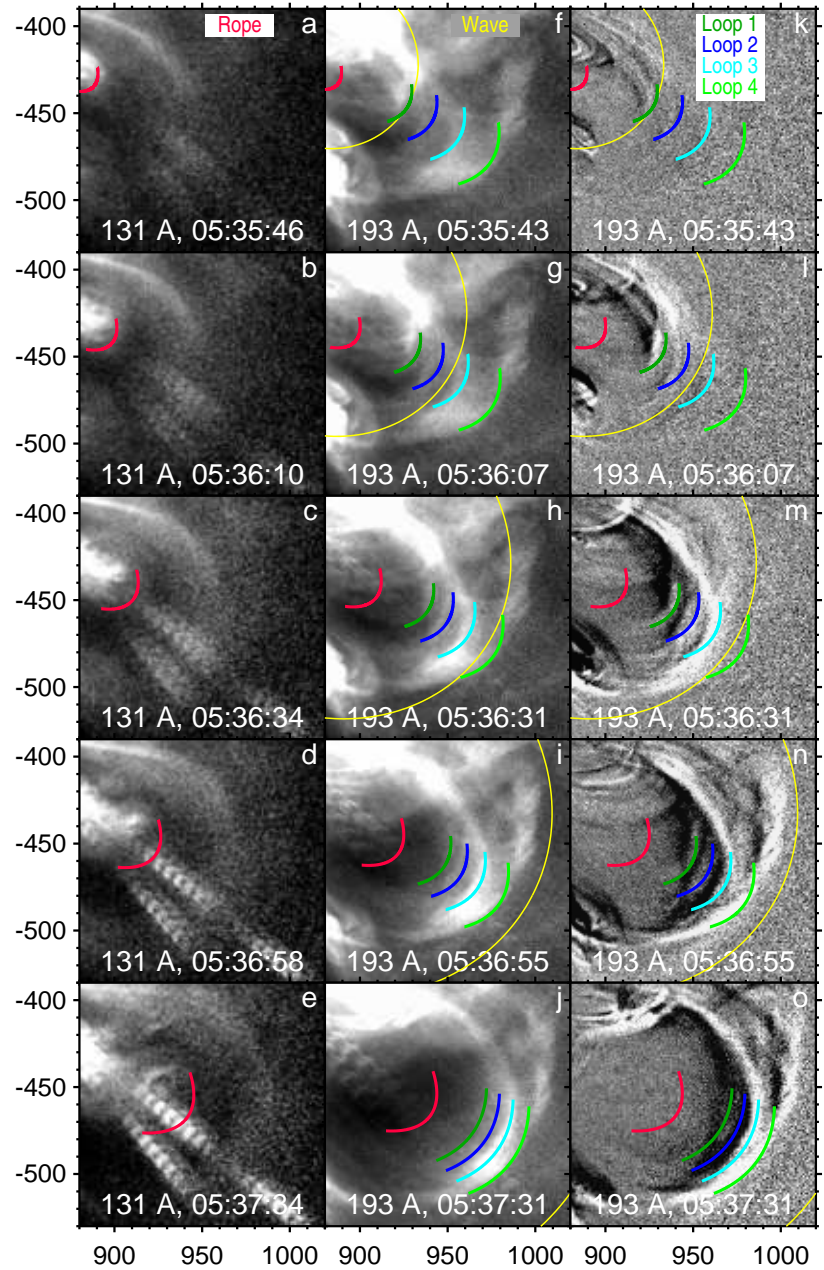}
  }
  \caption{The eruptive flux rope in AIA 131~\AA\ images
(left column, a--e) and the loops sequentially involved in the
eruption observed in 193~\AA\ (middle and right columns). The
middle column (f--j) presents the 193~\AA\ images with a reduced
background. The right column (k--o) presents their 12-s running
differences. The red arc outlines the top of the eruptive flux
rope. The yellow arc outlines the wave. The four other color arcs
approximately outline four conspicuous loops 1--4 of a
pre-eruption arcade.}
  \label{F-CME_formation}
  \end{figure}

Figure~\ref{F-CME_formation} and the \url{AIA_131_171_loops.mpg}
movie present some episodes of the CME genesis. The erupting flux
rope visible in 131~\AA\ is shown in the left column. The red arc
outlines its top according to the measurements in
Figure~\ref{F-rope_measurements}. The arc is also superimposed on
the 193~\AA\ images in the middle and right columns, where the
flux rope is not visible. The middle column with the
reduced-background 193~\AA\ images
(Section~\ref{S-image_processing}) presents the pre-eruptive
coronal arcade above the AR. Four individual loops 1--4 are
approximately outlined by the color oval arcs. The right column
shows a selection of 12~s running difference 193~\AA\ images. The
loops appear in these images, when start moving.

The CME formation process presented in the right column of
Figure~\ref{F-CME_formation} developed from below upwards, as the
preceding studies concluded. This succession is confirmed by the
progressively decreasing distances between loops 1--4. The agent,
which drove the loops, was an outward-propagating MHD disturbance.
The distance between the red and dark-green arcs in the middle
column (\textit{e.g.}, Figures \ref{F-CME_formation}f and
\ref{F-CME_formation}i) also decreases; the flux rope (red)
started to expand earlier and faster than the lowest loop~1
(dark-green). Thus, a probable driver of the expansion process
forming the CME was the flux rope, which erupted at a very small
altitude (left column).

The disturbance produced by the impulsively erupting flux rope is
represented by the yellow circle. After the passage of this
disturbance through loops 1 to 4, they sequentially start moving
(Figures \ref{F-CME_formation}k--\ref{F-CME_formation}n). The
loops become compressed to each other from below in Figures
\ref{F-CME_formation}j and \ref{F-CME_formation}o.

\begin{figure} 
  \centerline{\includegraphics[width=\textwidth]
   {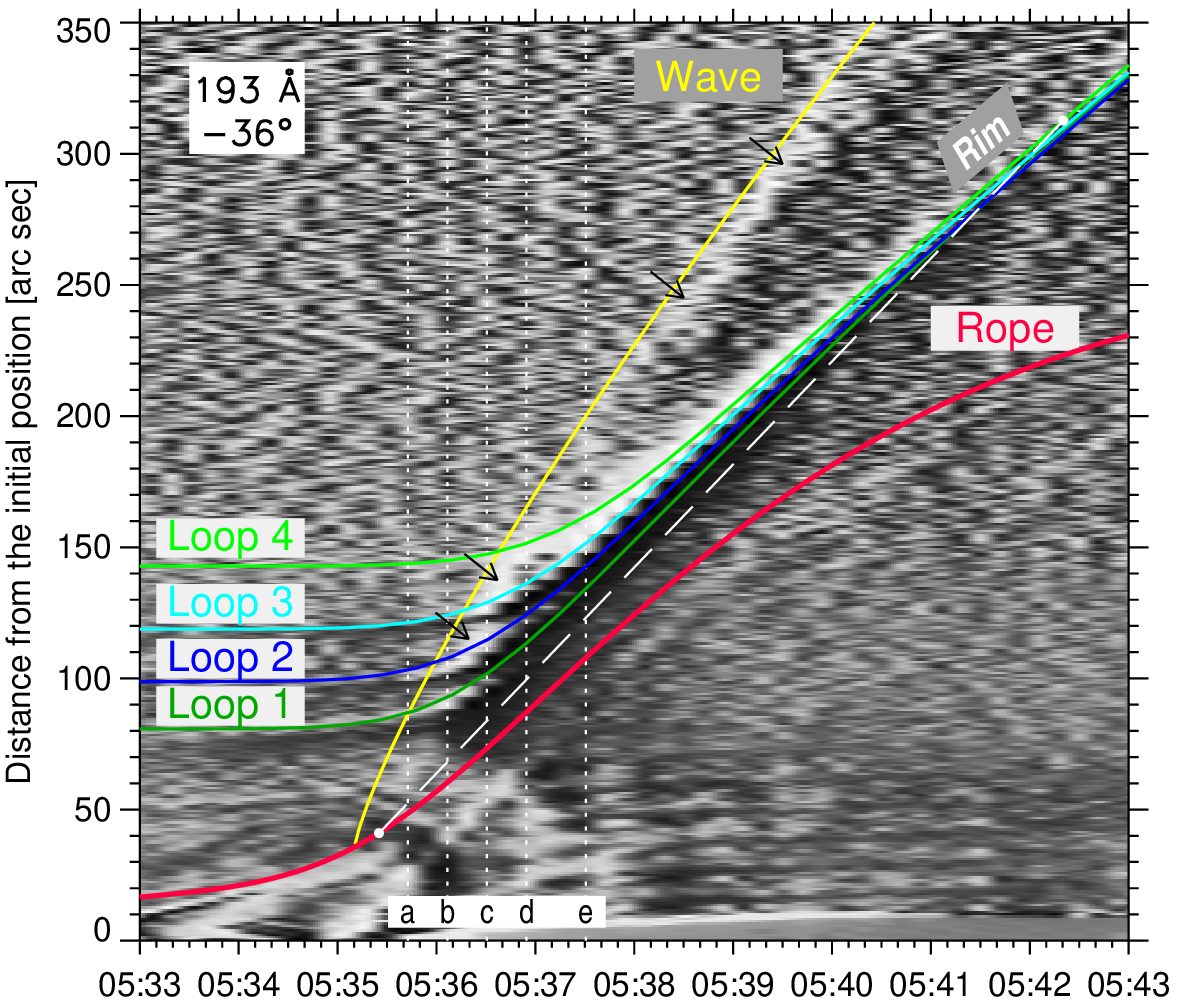}
  }
  \caption{CME formation in one-dimensional
spatial profiles computed from running-difference 193~\AA\ images
in a direction of $-36^{\circ}$ southward from the west. The red
curve represents the flux rope. The yellow curve outlines the
trace of the wave. Some of its signatures are indicated by the
arrows. The remaining color curves outline the trajectories of the
expanding arcade loops (same colors as in
Figure~\ref{F-CME_formation}). The vertical lines mark the
observation times of Figures
\ref{F-CME_formation}a--\ref{F-CME_formation}e. The labels of the
corresponding panels are indicated at the bottom. The tilted
dashed line denotes the trajectory of a virtual piston discussed
in Section~\ref{S-driver_CME_wave}.}
  \label{F-history_36_deg}
  \end{figure}

The outward-propagating disturbance and the involvement of the
loops in the expansion is demonstrated by the time history of the
CME formation in one-dimensional spatial profiles in
Figure~\ref{F-history_36_deg}. The profiles were computed from the
running-difference 193~\AA\ images in a fixed direction of
$-36^{\circ}$ southward from the west and averaged over a
10-pixels wide slice. The image in Figure~\ref{F-history_36_deg}
is similar to the slit images presented by \inlinecite{Ma2011} in
their Figure~3 but shows more details due to a harder image
processing.

Any moving feature appears here as an inclined strip, whose
instant slope is its velocity. The traces of the four loops 1--4
shown in Figure~\ref{F-CME_formation} are clearly visible.
Initially the loops rise slowly; in an interval of
05:35:40--05:37:30 their velocities considerably increase,
remaining nearly constant afterwards. Note that the running
differences only show the leading edge, while the trailing part
disappears.

The red curve is the flux-rope plot from
Figure~\ref{F-rope_measurements}b. A faint trace is detectable
$(15-20)^{\prime \prime}$ below it up to 05:38:30 and, possibly,
later. This trace seems to belong to the flux rope, while the lag
is partly due to the varying direction of its fastest expansion,
mostly different from $-36^{\circ}$ (see Figures
\ref{F-flux_rope}f--\ref{F-flux_rope}k). The presence of this
trace in 193~\AA\ (and 131~\AA) without any manifestations in 171
or 211~\AA\ indicates a temperature around 17~MK in the detected
part of the flux rope.

A fastest faint trace is detectable in the nearly radial direction
after 05:36:10 even inside the forming CME, starting at a height of
$\approx 95$~Mm. Its yellow outline was calculated for the wave
propagation using Equation~(\ref{E-pl_fit}) in
Section~\ref{S-kinematic_measurements} with a wave onset time
estimated for the flux rope's acceleration peak, $t_0 =
$~05:35:10. A density falloff exponent, $\delta = 2.5$, was adjusted
to reach a best fit of the wave trace. The kinematical identity of
the wave traces inside the rim and outside it rules out the
bow-shock excitation by the rim. The yellow ovals in
Figure~\ref{F-CME_formation} also correspond to this fit.
Although the power-law fit was derived for a blast wave, the wave
inside the forming CME, most likely, had not yet steepened into the
shock (to be discussed in Section~\ref{S-driver_CME_wave}).

The kinematics of loops 1--4 in Figure~\ref{F-history_36_deg} can
be inferred, keeping in mind that their expansion was limited from
above by the rim. Initially the loops were static. The
outward-propagating wave reached loops 1, 2, 3, and 4 one after
another, and sequentially drove their expansion. Loop~1 acquired a
highest speed and then had to decelerate, being restricted by the
rim. The highest speeds, accelerations, and decelerations of loops
2, 3, and 4 slightly decreased one after another. The final speeds
of the four loops converged to the final speed of the rim.

Four color curves outlining the trajectories of loops 1--4 were
calculated analytically with parameters adjusted to match the
traces of the loops. \inlinecite{Eselevich2013} presented
approximate height--time plots of the rising loops and proposed a
fast emergence of a magnetic tube from below the photosphere at
about 05:33. This idea contradicts the slow rise of the
CME-progenitor coronal structures during the long-lasting
pre-eruption heating and the early onset, by 05:31, of a sharp
increase in the 131~\AA\ and 0.5--4~\AA\ emissions in Figures
\ref{F-flux_rope}a--\ref{F-flux_rope}e and
Figure~\ref{F-rope_measurements}. Nevertheless,
\inlinecite{Eselevich2013} correctly showed in their Figure~7a the
sequential involvement of the loops in the motion.

\begin{figure} 
  \centerline{\includegraphics[width=0.8\textwidth]
   {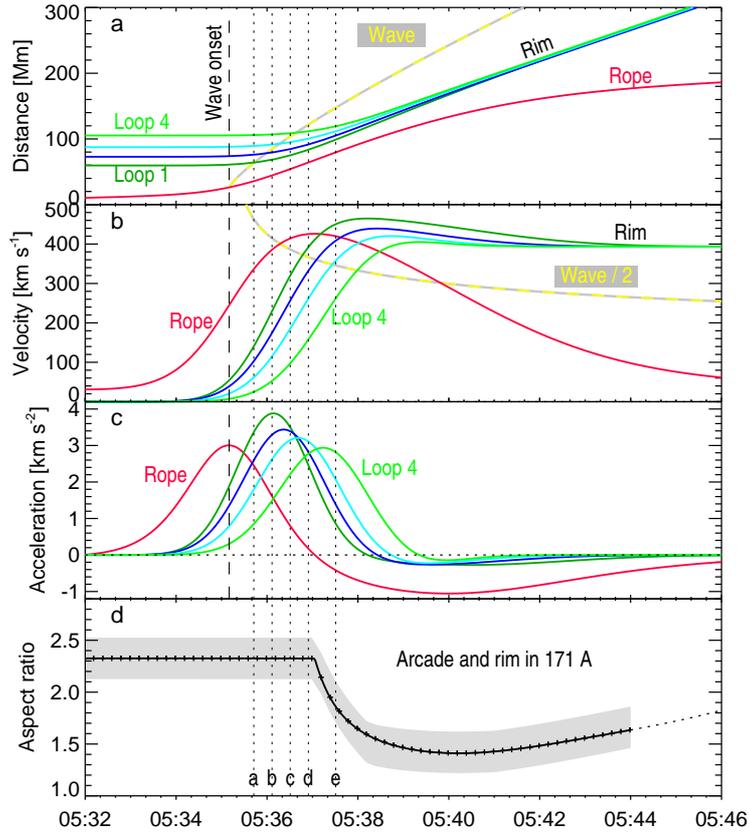}
  }
  \caption{Kinematics of the flux rope, arcade loops 1 -- 4,
and the wave (same colors as in Figures \ref{F-CME_formation} and
\ref{F-history_36_deg}). (a)~Distance--time plots similar to
Figure~\ref{F-history_36_deg}. The vertical dotted lines mark the
observation times of Figures \ref{F-CME_formation}a --
\ref{F-CME_formation}e. The labels of the corresponding panels are
indicated at the bottom. The vertical dashed line marks the wave
onset time. (b)~Velocity--time plots. The wave velocity (dashed
yellow-gray) in the plot is reduced by a factor of 2.
(c)~Accelerations of the flux rope (red) and arcade loops 1 -- 4.
(d)~The measured aspect ratio of the arcade loop~4 and rim. The
shading represents the uncertainty. The dotted extension
approximately corresponds to the later expansion, when the rim is
not reliably detectable.}
  \label{F-kinematics_loops}
  \end{figure}

The analytic color height--time plots in
Figure~\ref{F-history_36_deg} allowed us to obtain quantitative
kinematics of the loops. The corresponding plots for the four
loops, flux rope (red) and wave (dashed yellow-gray) are presented
in Figures \ref{F-kinematics_loops}a--\ref{F-kinematics_loops}c.

Figure~\ref{F-kinematics_loops}a reproduces
Figure~\ref{F-history_36_deg} without the background image.
Figure~\ref{F-kinematics_loops}b shows the velocity--time plots.
The highest velocities reached by the flux rope and the four loops
were between 400 and 500 km~s$^{-1}$. The wave speed was actually
twice higher than the dashed yellow-gray curve shows, and started
from $\gsim 1000$~km~s$^{-1}$, which is a typical fast-mode speed
in the low corona above an active region. Then the wave speed
monotonically decreased all the time.

The flux rope speed was $\approx 250$~km~s$^{-1}$ at the wave
onset time, $t_0 = $~05:35:10, and rose farther. Loop~4 and the
forming rim started to expand well after the wave onset time, and
therefore could not excite the wave. The relation between the
velocities of the flux rope and wave rules out its bow-shock
regime. Note that the loops accelerated gradually which indicates
that the wave was not yet in the shock regime until, at least, its
passage through loop~4 at 05:36:30 in the direction of the
measurements $-36^{\circ}$ southward from the west. Otherwise, the
velocity of a loop pushed by a shock wave should change abruptly.

Figure~\ref{F-kinematics_loops}c presents the accelerations of the
flux rope and loops. All components of the forming CME were
obviously driven by the erupting flux rope, whose acceleration
pulse led all others. Loops 1--4 sequentially accelerated
($3-4$~km~s$^{-2}$) up to $400-500$~km~s$^{-1}$, and then somewhat
decelerated, approaching the final speed of the rim. A trivial
deceleration of the wave is not shown.

Figure~\ref{F-kinematics_loops}d shows the aspect ratio estimated
for the top of the forming CME in manual outlining the curvature
of loop~4, which then joined the rim (the black arc in
Figure~\ref{F-bubble171}). The uncertainty is shown by the
shading. We measured the aspect ratio by outlining loop~4
with an ellipse, whose lower edge was fixed at the photosphere,
and did not endeavor to catch the whole shape of the loop, which
was more complex. The aspect ratio variations are best visible in
the \url{AIA_131_171_loops.mpg} movie, where loop~4 and the rim
are outlined by the green arc. Comparison of Figures
\ref{F-kinematics_loops}d and \ref{F-kinematics_loops}c reveals a
similarity between the variations in the aspect ratio and the
flux-rope deceleration. However, the aspect ratio increased slower
than the deceleration ceased (\textit{cf.} the shapes of the two
curves after 05:40). Thus, the variations in the aspect ratio of
the CME bubble were probably governed by the flux rope expanding
inside it, while its reaction had a reasonable delay.

It is worth to compare our measurements with the results obtained
previously. \inlinecite{Patsourakos2010} were the first who
measured the speed, acceleration, and the aspect ratio of the rim
(CME bubble). They found that its speed reached a maximum of
400~km~s$^{-1}$ at 05:38 and then decreased to 300~km~s$^{-1}$.
The maximum acceleration of 2~km~s$^{-2}$ was found to occur
slightly after 05:36, followed by a deceleration up to
$-0.5$~km~s$^{-2}$ around 05:39. With quite different measurement
techniques used by us and the authors, both results appear to be
close to each other with an acceptable accuracy.

The aspect ratio variations we measured are basically close to
those found by \inlinecite{Patsourakos2010}, with somewhat larger
differences due to different ways of the measurements. Substantial
is a recovery tendency of the aspect ratio after 05:40 which is
also indicated by the measurements of \inlinecite{Gopalswamy2012}
in their Figure~3b. Possible causes of these variations are
discussed in Section~\ref{S-rim}.

The difficulties to detect the earliest signatures of the wave and
to measure its kinematics have resulted in a scatter between the
results of different authors. \inlinecite{Kozarev2011} found the
wave speed of $\approx 735$~km~s$^{-1}$ at 05:37 and its
subsequent deceleration, very close to our results. The estimates
by \inlinecite{Ma2011} show a certain deceleration after 05:40
from 600~km~s$^{-1}$ to about 500~km~s$^{-1}$ at 05:42--05:44,
also mainly consistent with our results.

\subsubsection{Rim and Pileup on its Top in 211~\AA\ AIA Images}

\begin{figure} 
  \centerline{\includegraphics[width=\textwidth]
   {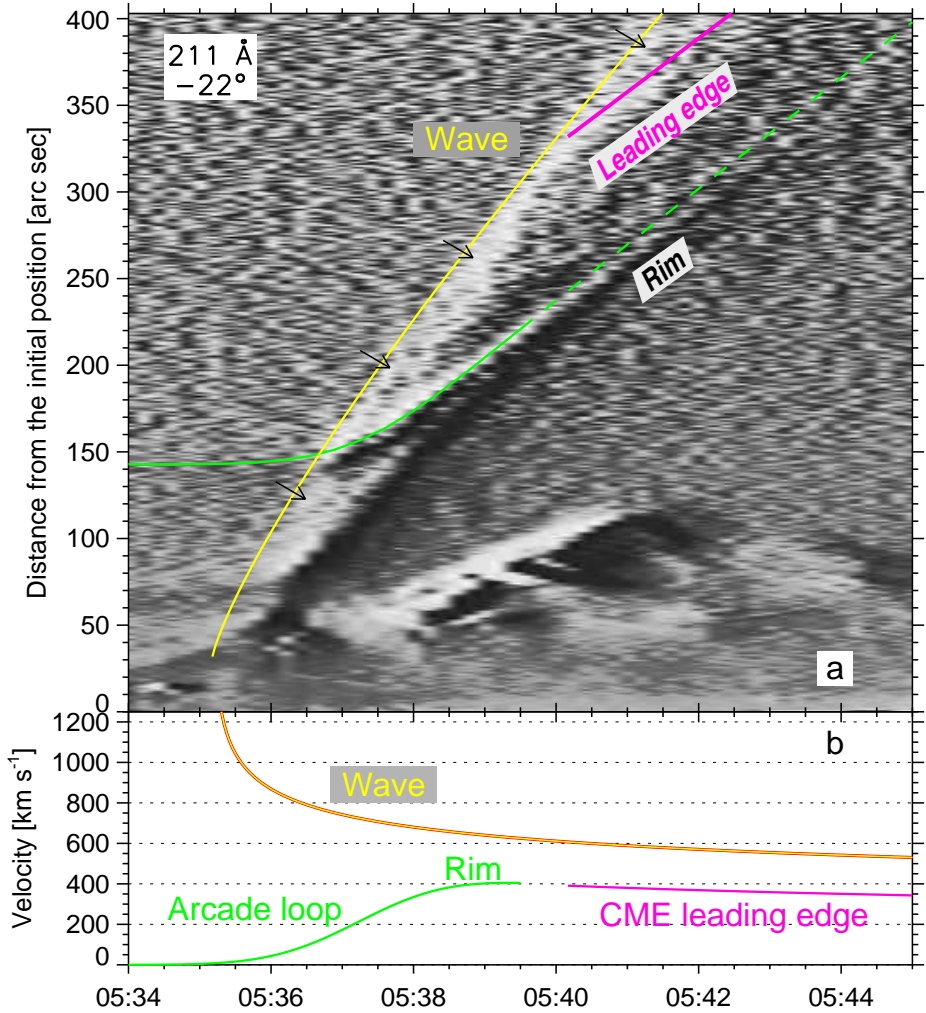}
  }
  \caption{(a)~Time history of the CME formation in one-dimensional
spatial profiles computed from running-difference AIA 211~\AA\
images in a direction of $-22^{\circ}$ southward from the west.
The yellow curve outlines the wave trace. Some of its signatures
are indicated by the arrows. The green curve corresponds to the
arcade top (loop~4) clinging to the rim. The later part of the rim
signature shows a stronger deceleration then loop~4 in Figures
\ref{F-history_36_deg} and \ref{F-kinematics_loops} (the dashed
continuation of the green curve) had. The outermost signature of
the bubble is outlined with the pink curve corresponding to the
leading edge of the CME frontal structure (the top of piled-up
plasma). (b)~Velocity--time plots for the wave (yellow on gray),
the upper arcade loop 4 and the rim (green), and the CME leading
edge (pink).}
  \label{F-history_22_deg}
  \end{figure}

To study the details of the pileup formation on top of the rim
shown in Figure~\ref{F-bubble211}, we consider the 211~\AA\ AIA
images in a way similar to the preceding section. Here we use a
different direction, in which the loops inside the rim are
indistinct, but the rim, pileup, and wave trace are clearly
visible. Figure~\ref{F-history_22_deg}a presents a time-history
diagram computed from running differences of the 211~\AA\ images
in a direction of $-22^{\circ}$ southward from the west.

The trace of the arcade top represented by loop~4, which joined
the rim, is distinct by about 05:40, and later it becomes poorly
visible. Nevertheless, comparison of its faint trace with the
dashed continuation of the earlier trajectory measured from
Figure~\ref{F-history_36_deg} indicates that the rim starts
decelerating.

A much faster wave trace is outlined by the yellow fit. The
211~\AA\ data reveal an additional, slower bright branch outlined
by a pink curve. It goes nearly parallel to the trace of the rim,
being $\approx 90^{\prime \prime}$ higher. The high trace
corresponds to the white arc in Figure~\ref{F-bubble211} outlining
the outer edge of the pileup. To relate it to the white-light CME,
we calculated the kinematics of the frontal structure using the
self-similar approximation \cite{Grechnev2014_II} with the
parameters adjusted to coordinate the AIA and SOHO/LASCO
observations. The parameters of the pink outline are $V_1 =
380$~km~s$^{-1}$ and $r_1 = 264$~km ($360^{\prime \prime}$) at
$t_1 =$~05:41:00, and $V_\infty = 190$~km~s$^{-1}$ (the solid line
in Figure~\ref{F-lasco_kinematics}). The compression of the loops
to the rim indicates that the self-similar approximation did not
yet fully apply at 05:41:00 (otherwise, the CME would expand
uniformly); nevertheless, the pink outlines in Figures
\ref{F-history_22_deg}a and \ref{F-history_22_deg}b appear to be
acceptable. Our technique does not yet allow smooth concatenating
the green rim's velocity with the pink CME speed, although they
are close to each other.

\begin{figure} 
  \centerline{\includegraphics[width=\textwidth]
   {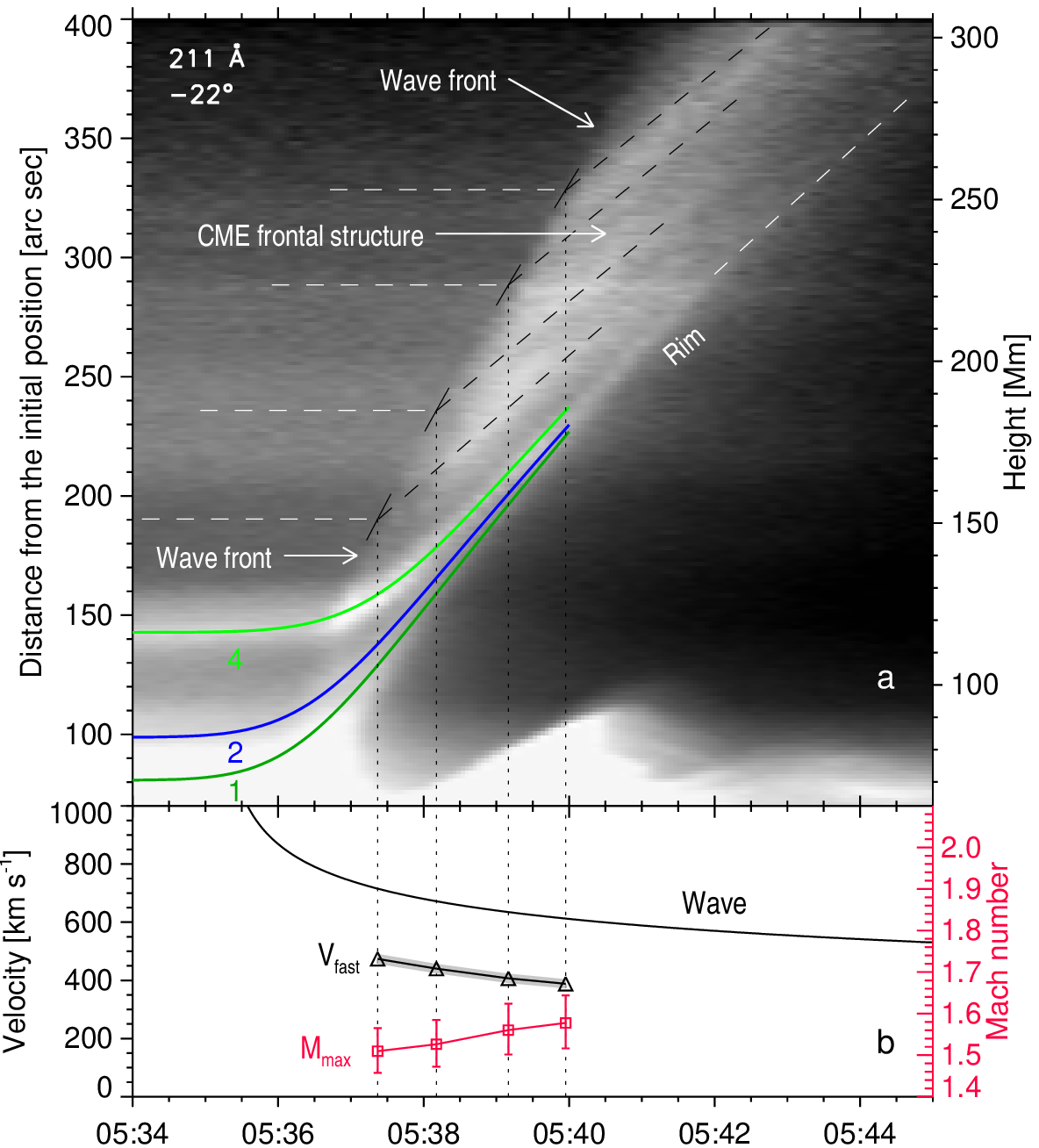}
  }
  \caption{(a)~Formation of the CME frontal structure in
spatial profiles computed from 211~\AA\ fixed-base differences in
a direction of $-22^{\circ}$ southward from the west. The black
dashed lines outline individual structures involved in the motion
by the wave, whose front is outlined by the rare dashes. The white
horizontal dashed lines left of the wave front denote presumable
initial positions of these structures. The color curves labeled 1,
2, and 4 correspond to the loops discussed previously. The
inclined white dashed line outlines the rim. (b)~Velocity--time
plots for the wave (long black curve), the maximum Mach number
(red squares) with uncertainties, and the calculated fast-mode
speed (triangles) with uncertainties shown by the shading.}
  \label{F-history_fs}
  \end{figure}

To understand the pileup formation better and to figure out the
properties of the wave, we consider a similar diagram in
Figure~\ref{F-history_fs} computed in the same direction from
fixed-base difference 211~\AA\ images. It was very difficult to
reveal individual structures between the rim and the leading edge,
and therefore their separate traces outlined by the black dashed
lines are regrettably faint. Several attempts showed that the
errors in estimating the slopes of the faint traces in
Figure~\ref{F-history_fs}a (\textit{i.e.}, their velocities) did
not exceed $\pm 7\%$.

Previous studies assumed that the shock formed before 05:37, when
the type~II burst started, but it is unknown when the wave entered
the shock regime in the direction of $-22^{\circ}$ in
Figure~\ref{F-history_fs}. One cannot recognize if the transition
between the horizontal dashed lines left and right from the wave
front was abrupt (shock) or gradual. For the Mach number, $M$, it
is only possible to estimate the upper limit, $M_{\max} =
V_{\mathrm{sh}}/V_{\mathrm{fast}}$, so that $1 \leq M \leq
M_{\max}$.

With known velocities of the wave front, presumably shock,
$V_{\mathrm{sh}}$, and a structure moved by the gas behind the
shock, $U_{\mathrm{sh}}$, one can estimate the fast-mode speed,
$V_{\mathrm{fast}}$, from an equation $V_{\mathrm{sh}} \approx
V_{\mathrm{fast}} + \kappa U_{\mathrm{sh}}/2$; the $\kappa$
coefficient governs the wave steepening rate
\cite{Grechnev2011_I}. This coefficient, $1/2 \leq \kappa \leq
3/2$, depends on plasma beta and the propagation direction
(\opencite{AfanasyevUralov2012}, Figure~8). Most likely, here we
are dealing with a wave propagation nearly perpendicular to the
magnetic field in low-beta plasma, $\kappa \approx 3/2$. The
estimates of the maximum Mach number and the fast-mode speed for
four instants are shown in Figure~\ref{F-history_fs}b. Such
estimations do not depend on the blast-wave or bow-shock regime,
while their accuracy decreases for strong shocks.

Since the trajectory of loop~4 in Figure~\ref{F-history_fs}a is
gradual, the discontinuity had not yet formed at 05:36:45.
Figure~\ref{F-history_fs}b indicates that the Mach number could
only increase from 05:37:25 to 05:40:00, not exceeding $M_{\max}$,
which was nearly constant, $1.45 \leq M_{\max} \leq 1.65$. The
wave probably evolved from a linear fast-mode wave ($M = 1$) to a
simple wave and then steepened in some time into the shock with $M
< M_{\max}$. The shock might have not formed at all in this
direction, right ahead of the CME. The source of the type~II burst
could be located at a flank of the wave front---\textit{e.g.}, in
a streamer-like structure in Figure~\ref{F-bubble211}a.

The fast-mode speed shown in Figure~\ref{F-history_fs}b was
calculated under an assumption of the weak shock regime of the
wave. If this was not the case for all the four instants, then the
real fast-mode speed should be between the line connecting the
triangles and the higher plot of the wave, anyway decreasing with
height. The decrease is typical of the fast-mode speed above
active regions at heights $< 0.4R_{\odot}$ \cite{DulkMcLean1978,
Gary2001, Mann2003}.

In summary, Figures \ref{F-history_22_deg} and \ref{F-history_fs}
demonstrate that CME formed due to the outward-propagating wave,
which swept up all the structures in its way, involving them in
the expansion. The CME frontal structure was mostly constituted by
the pileup on top of the expanding rim that previously was a
relatively high environment of the active region.

\subsection{CME Expansion Visible in White Light}
 \label{S-lasco}

\begin{figure} 
  \centerline{\includegraphics[width=\textwidth]
   {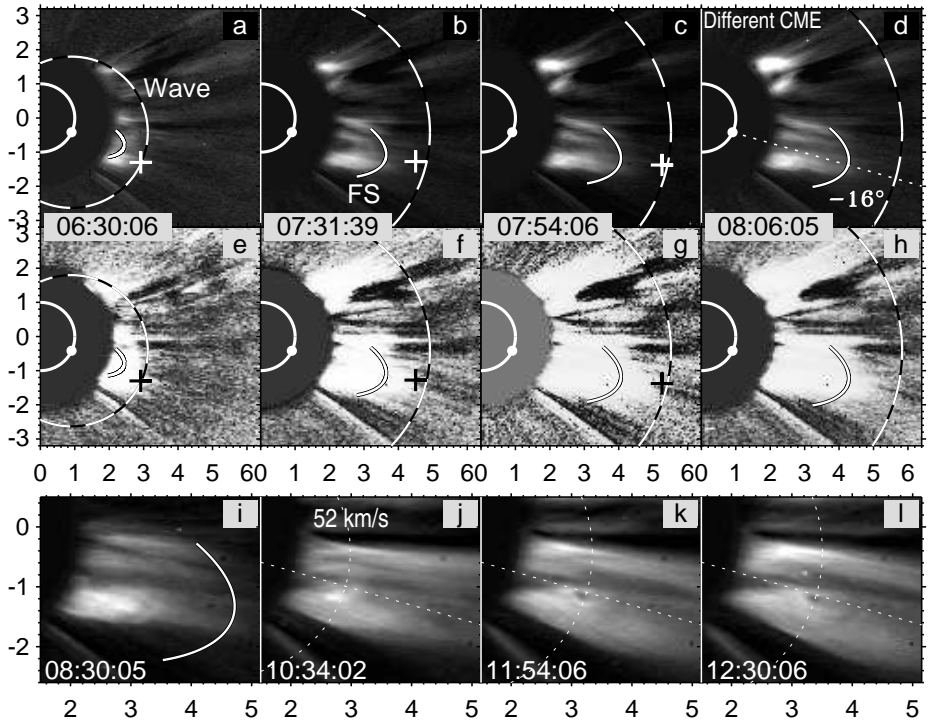}
  }
  \caption{LASCO/C2 images: CME expansion (upper row, fixed ratios),
wave traces (middle row, running differences), and the CME
structure (lower row, fixed ratios). The small oval arcs outline
the leading edge of the frontal structure (FS). The larger dashed
circles in the upper and middle rows outline the wave traces. The
crosses denote the measurements in the CME catalog. The dotted
circle in the lower row corresponds to the velocity of the flux
rope's center of $\approx 52$~km~s$^{-1}$. The solid circle
denotes the solar limb. The small filled circle denotes the
eruption site. The axes present the coordinates from the solar
disk center in solar radii.}
  \label{F-lasco}
  \end{figure}

\begin{figure} 
  \centerline{\includegraphics[width=0.8\textwidth]
   {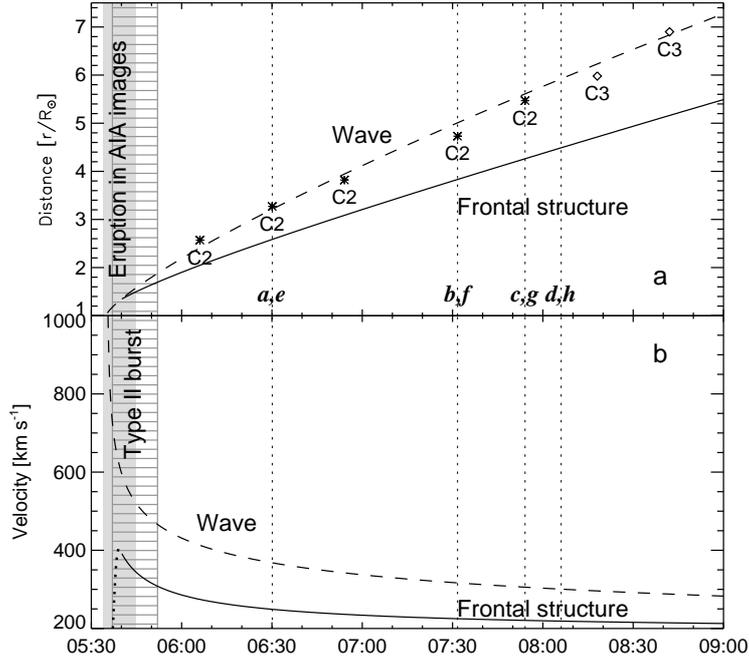}
  }
  \caption{Heliocentric distance--time (a) and velocity--time (b) plots
for the leading edge of the CME frontal structure (solid) and the
wave (dashed). The symbols in panel (a) represent the measurements
from the CME catalog. The vertical dotted lines denote the times
of the images in Figure~\ref{F-lasco}. The labels of the
corresponding panels are indicated at the bottom of panel (a). The
initial accelerating part (dotted) in panel (b) corresponds to the
arcade loop and the rim in Figure~\ref{F-history_22_deg}. The gray
shading corresponds to the time interval presented in
Figure~\ref{F-history_22_deg}. The horizontal hatching denotes the
interval, in which the type II burst was observed.}
  \label{F-lasco_kinematics}
  \end{figure}

To validate our results drawn from the EUV observations of the
eruption, it is important to coordinate them with a white-light
CME observed by SOHO/LASCO. Figure~\ref{F-lasco} shows selected
LASCO/C2 images. The CME leading edge and wave traces are outlined
according to the kinematics presented in
Figure~\ref{F-lasco_kinematics}.

The upper and middle rows of Figure~\ref{F-lasco} shows different
representations of the same four images to reveal a poorly visible
CME structure in the upper row and to detect faint wave traces in
the middle row. The faint frontal structure (FS) is outlined by
the oval arc, whose increasing radius corresponds to the pink
self-similar fit in Figure~\ref{F-history_22_deg}. It seems to
consist of stretched loops. The rim is not pronounced in the CME
structure. The CME orientation still turned from the initial
$-46^{\circ}$ in Figure~\ref{F-flux_rope}f to $-16^{\circ}$ in
Figure~\ref{F-lasco}d (position angle of $254^{\circ}$ in the CME
catalog \url{http://cdaw.gsfc.nasa.gov/CME_list/};
\opencite{Yashiro2004}).

The crosses in the upper and middle rows represent the
measurements in the CME catalog. They were made for the fastest
feature, being close to the wave traces outlined by the dashed
circle. The wave outline is the same as we used in the preceding
section, with $t_0=$~05:35:10 and $\delta = 2.5$.

The four later C2 images in the lower row reveal a flux-rope
structure of the CME core. The average speed of its center marked
with the dotted circle in Figures \ref{F-lasco}j--\ref{F-lasco}l
is 52~km~s$^{-1}$, consistent with our measurements in
Figure~\ref{F-rope_measurements}c. The flux rope, whose initial
expansion drove the whole CME formation process, later relaxed and
became the CME core visible well behind the leading edge.

The kinematic evolution of the FS and the wave ahead it is clear
from Figure~\ref{F-lasco_kinematics}. The FS was probably formed
from coronal loops swept up by the expanding rim, whose velocity
is plotted in Figure~\ref{F-lasco_kinematics}b with the dotted
line. Being expelled by the erupting flux rope, the wave initially
was fast and possibly strong enough to produce the type II
emission within the hatched interval. The wave speed in the radial
direction decreased within this interval from $\approx 680$ to
$\approx 460$~km~s$^{-1}$. Note that the type II burst source
could be located in a different direction, where the wave strength
might be also different.

Then the wave strongly decelerated and dampened, being not driven
by the trailing piston, which considerably slowed down. The
evolution of the wave speed inferred from AIA and LASCO
observations does not confirm an assumption of some authors about
its possible peak between $1.5R_{\odot}$ and $2.6R_{\odot}$ (the
appearance in the LASCO field of view). Although the wave and CME
were kinematically similar, the wave speed at distances $>
2R_{\odot}$ was too low for the bow-shock regime. Then the wave
speed still decreased below 300~km~s$^{-1}$ at about $7R_{\odot}$,
comparable to the solar wind speed, that points to its decay into
a weak disturbance. The increasing role of the solar wind is
confirmed by a subsequent acceleration of the CME suggested by the
measurements in the CME catalog.

\section{Wave Signatures and EUV Transient}
 \label{S-Wave}

\subsection{EUV Wave}
 \label{S-EUV_wave}

Most preceding studies considered a transient expanding in EUV
images (EUV wave; \opencite{Warmuth2015}) as a signature of a
shock wave, assuming its bow-shock regime. We showed in the
preceding sections that the EUV wave moving away from the Sun
consisted of swept-up plasmas on top of the expanding separatrix
surface, with which the rim was associated. The pileup was
involved in the motion by the outward-propagating wave, which was
initially excited by the impulsive expansion of the flux rope, and
then resembled decelerating blast wave.

The propagation conditions along the solar surface are
considerably different from those away from the Sun. In the
lateral directions, the separatrix surface does not follow the
expanding wave up to large distances. It is therefore important to
study the EUV wave propagating in different directions.

\inlinecite{Downs2012} analyzed the EUV wave in this event in
realistic coronal conditions on the basis of a thermodynamic MHD
simulation. They stated a clear distinction between the wave and
non-wave component and concluded that the propagating EUV
transient exhibited the behavior of a fast-mode wave. However, it
was difficult to ascertain the wave excitation scenario in this
simulation.

Conversely, our approach only allows studying global properties of
a wave, without a reference to realistic inhomogeneous corona. We
will nevertheless try to reconcile the wave excitation scenario
revealed with a posterior near-surface wave propagation and
examine how acceptable its fit in this case works.

\begin{figure} 
  \centerline{\includegraphics[width=\textwidth]
   {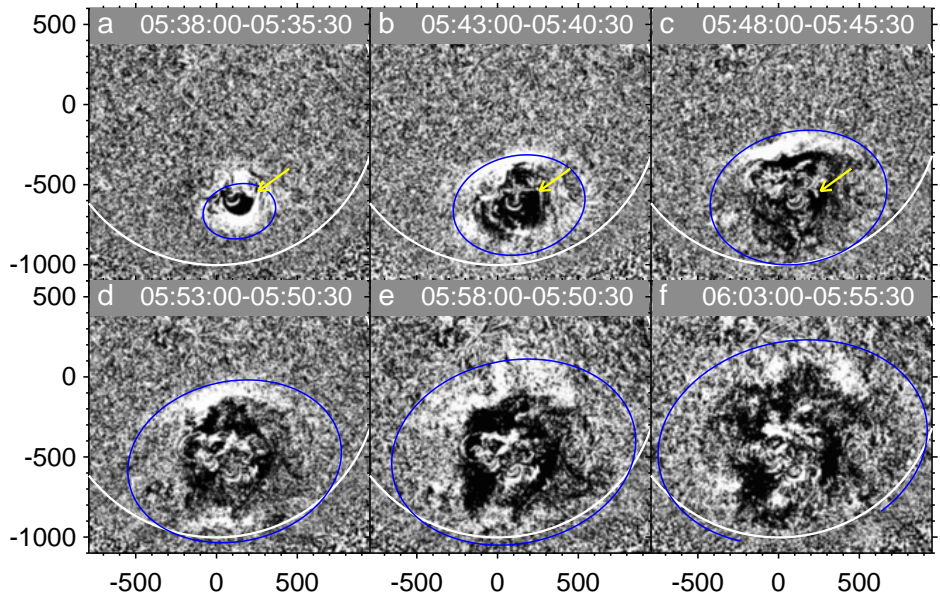}
  }
  \caption{EUV wave propagation in STEREO-A/EUVI 195~\AA\ running difference
images. The blue ellipses represent calculated wave fronts. The
yellow arrow in panels (a)--(c) points at a base of the
streamer-like feature (denoted in Figure~\ref{F-bubble211}a), in
which the source of the type II burst could be located. The white
circles outline the solar limb. The axes show the coordinates in
arcsec from the solar disk center.}
  \label{F-euvi_wave}
  \end{figure}

\begin{figure} 
  \centerline{\includegraphics[width=0.7\textwidth]
   {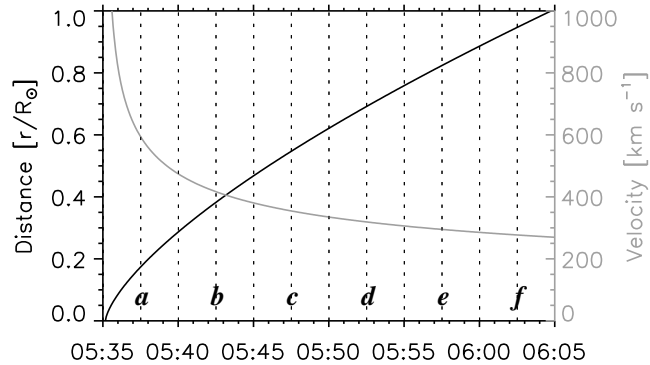}
  }
  \caption{Distance--time (black) and velocity--time (gray) plots of the
EUV wave propagation from the eruption center along the solar
surface. The vertical lines mark the observation times of Figures
\ref{F-euvi_wave}a -- \ref{F-euvi_wave}f corrected for the
difference between the orbits of the Earth and STEREO-A, as if the
Sun were viewed from SDO. The labels of the corresponding panels
are indicated at the bottom.}
  \label{F-euvi_wave_kinematics}
  \end{figure}

The EUV wave was observed from two vantage points by SDO/AIA and
STEREO-A. Some EUVI 195~\AA\ images are shown in
Figure~\ref{F-euvi_wave}, where their nominal observation times
are specified. The kinematical plots corresponding to the ellipses
outlining the EUV wave are presented in
Figure~\ref{F-euvi_wave_kinematics}, where the EUVI observation
times are referred to the SDO vantage point. The wave kinematics
along the spherical solar surface was calculated with the same
onset time as previously, $t_0 = $~05:35:10, and a density falloff
exponent of $\delta = 2.1$. The blue ellipses in
Figure~\ref{F-euvi_wave} (yellow in the \url{EUVI_wave.mpg} movie)
delineate the near-surface isotropic trail of an expanding global
wave front. This trail corresponds to an effective height of 35~Mm
and the wave propagation in the corona without any
inhomogeneities. The ellipses were calculated as small circles at
a sphere with a pole coinciding to the eruption center.

The ellipses tolerably correspond to the leading edge of the
bright EUV wave in the images, although the initial wave could
actually be somewhat faster than the outline. The effective height
might not be constant, and a small displacement of the wave
`epicenter' is not excluded. Such a displacement toward a region
of a higher fast-mode speed was reported previously (see,
\textit{e.g.}, \opencite{Grechnev2013_20061213}). The faintness of
the EUV wave here disfavors detection of this effect.

The most probable source of a type II radio burst is the current
sheet of a small coronal streamer stressed by a shock front (see
Section~\ref{S-type_II}). The type II burst in this event started at
05:37:00 (Figure~\ref{F-type_II_spectrum}). The yellow arrow in the
upper row of Figure~\ref{F-euvi_wave} points at the base of a
streamer-like feature visible in the AIA 211~\AA\ image in
Figure~\ref{F-bubble211}a. The wave front in
Figure~\ref{F-euvi_wave}a has already passed this feature, while the
observation time of this STEREO-A/EUVI image corresponds to
05:37:31~UT. Thus, the actual positions of the wave front do not
contradict a possible location of the type II burst source in this
feature.

\begin{figure} 
  \centerline{\includegraphics[width=\textwidth]
   {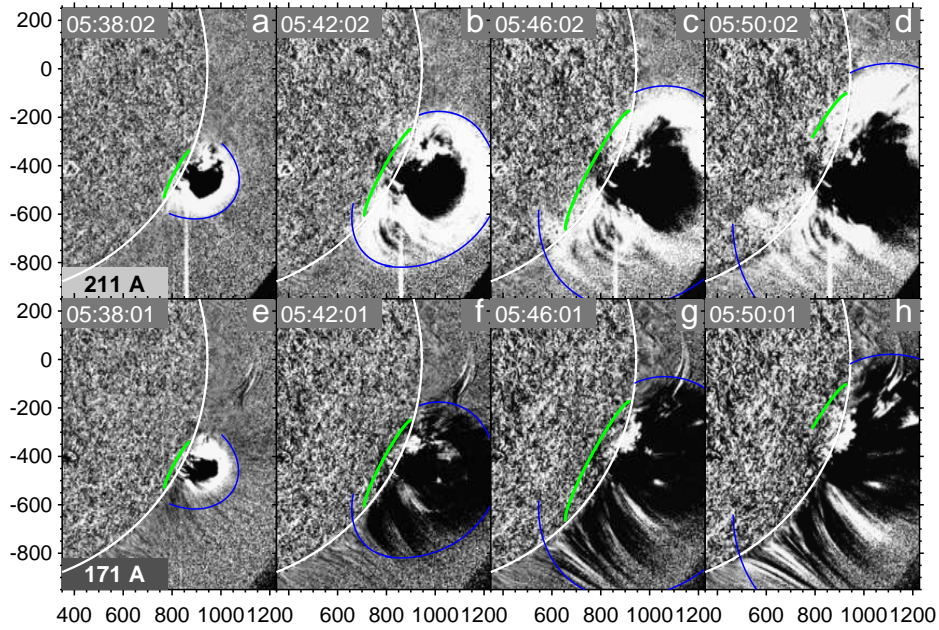}
  }
  \caption{EUV wave propagation in AIA 211 and 171~\AA\ fixed-base
image ratios to those at about 05:34:00. The blue curves outline the
off-limb wave front. The green ellipses outline its surface trail.
Two right panels reveal the wave reflection in the southern region.
The white circles trace the limb. The coordinates are in arcsec from
the solar disk center.}
  \label{F-aia211_wave}
  \end{figure}

Figure~\ref{F-aia211_wave} presents the EUV wave propagation as
seen by AIA in 211 and 171~\AA. The off-limb front is outlined by
the blue curves composed from three oval arcs adjusted to fit the
wave kinematics measured from the quadrature observations with
STEREO-A/EUVI and SDO/AIA. The orientations of the north and south
arcs were progressively adjusted to catch the tilt of the wave
front that apparently varied in its motion along the limb. The
radius of the south arc calculated from the power-law fit was
stretched by a constant factor to catch a faster wave propagation
toward the south pole, where the fast-mode speed was higher in the
region of the polar coronal hole. The green ellipses were
calculated for the surface trail of the wave front visible on the
Earth-facing hemisphere.

The EUV transient appears between the calculated wave front and
the rim as a brightening in the higher-temperature 211~\AA\ images
and as a darkening in the lower-temperature 171~\AA\ images. The
brightening in the first 171~\AA\ image in
Figure~\ref{F-aia211_wave}e is due to separate loops, which have
not yet merged into the thin rim. The EUV transient is most likely
due to the pileup, while the different appearance in the two
different-temperature channels indicates its heating.

\begin{figure} 
  \centerline{\includegraphics[width=\textwidth]
   {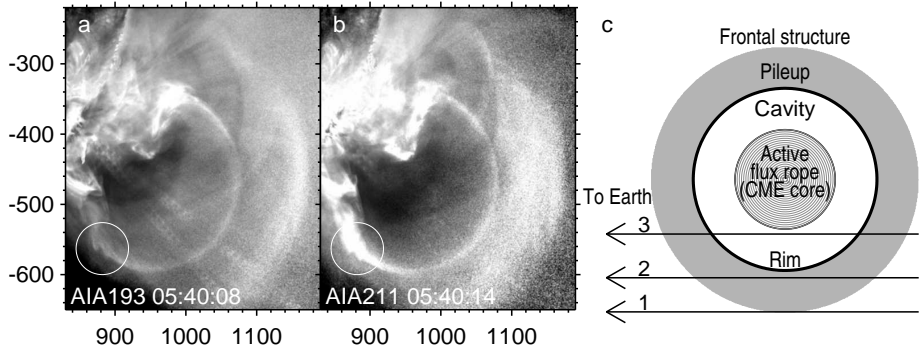}
  }
  \caption{CME bubble in the 193~\AA\ (a) and 211~\AA\ (b)
background-subtracted images at the indicated times divided by the
images averaged within 05:31:45--05:33:25. The circle outlines the
trace of a lateral disturbance, which follows the flux-rope liftoff
and runs along the separatrix. The sketch in panel (c) presents the
bubble as viewed approximately from STEREO-A and illustrates the
brightness distribution visible from SDO.}
  \label{F-scheme}
  \end{figure}

Figure~\ref{F-scheme} compares the appearance of the EUV transient
in the 193~\AA\ and 211~\AA\ images. They show a large difference
between the orientations of the rim (along with the arcade loops
pressed to the rim) and the long loops above it. This fact
corroborates the association of the rim with a separatrix surface.

Comparison of Figures \ref{F-scheme}a and \ref{F-scheme}b
indicates a larger opacity of the pileup on top of the rim in the
higher-temperature 211~\AA\ channel (excluding the hottest window
in 193~\AA) and some temperature increase from the outer edge of
the pileup to the rim. The appearance of the pileup suggests that
it was a thick, nearly spherical layer, bounded by the rim from
inside, as shown in Figure~\ref{F-scheme}c.

In summary, the EUV transient enveloping the rim was the pileup
and became the CME frontal structure. The near-surface EUV
transient observed at large distances in Figures
\ref{F-euvi_wave}b--\ref{F-euvi_wave}f and moving toward the south
pole in Figures \ref{F-aia211_wave}b--\ref{F-aia211_wave}c was a
trace of the wave, which was not followed by CME structures.
Having risen and expanded enough, the CME was not detectable from
STEREO-A against the solar surface due to a strong decrease of its
emission measure.

\subsection{Type II Burst}
 \label{S-type_II}

\begin{figure} 
  \centerline{\includegraphics[width=\textwidth]
   {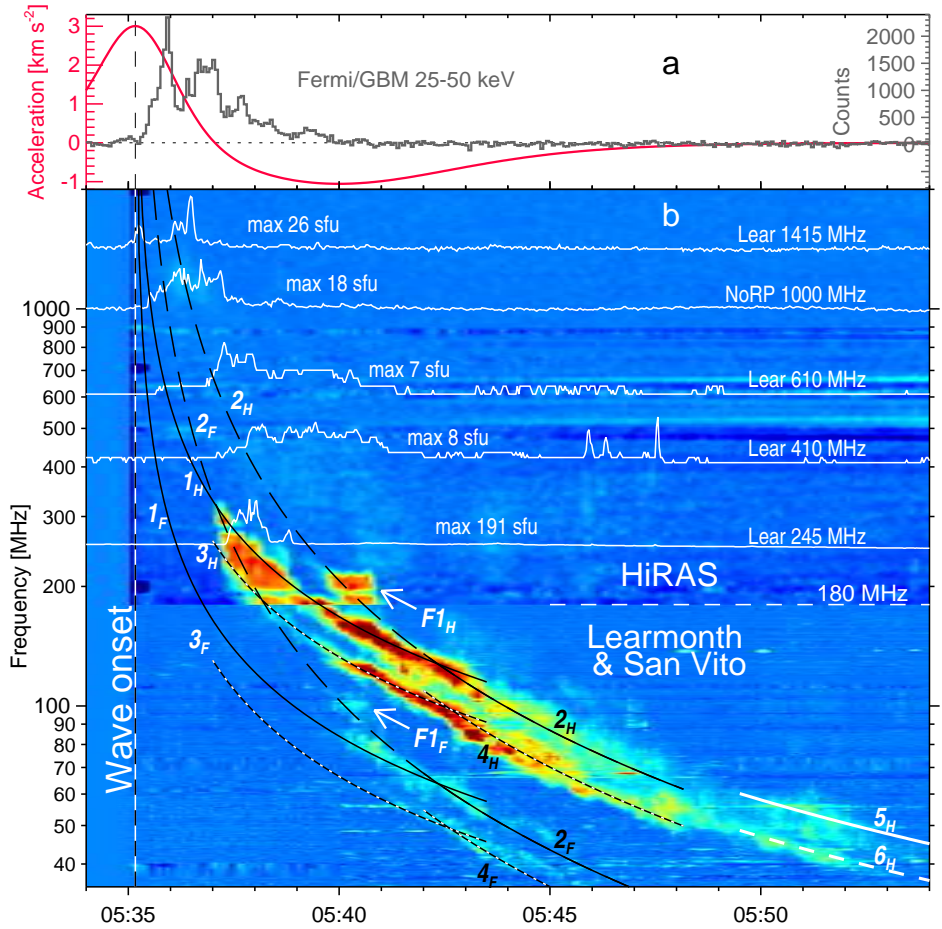}
  }
  \caption{Relations between the eruption, HXR burst,
and radio signatures of the shock wave. (a)~Acceleration of the
flux rope (red) and the HXR burst (gray). The vertical dashed line
marks the wave onset time. (b)~Dynamic spectrum of the type II
burst composed from the HiRAS data ($> 180$~MHz) and Learmonth and
San Vito RSTN data ($< 180$~MHz) along with normalized
fixed-frequency time profiles (Learmonth and NoRP). The
harmonically related pairs of the type II bands are outlined by
the calculated trajectories ($1_\mathrm{F}$, $1_\mathrm{H}$),
($2_\mathrm{F}$, $2_\mathrm{H}$), \textit{etc.} of different line
styles. An additional harmonic feature F$1_\mathrm{F}$,
F$1_\mathrm{H}$ is indicated by the arrows.}
  \label{F-type_II_spectrum}
  \end{figure}

The type II burst in this event was analyzed previously
\cite{Kozarev2011, Ma2011, Gopalswamy2012, Vasanth2014,
Kouloumvakos2014} based on data from different radio
spectrographs, each of which has its own advantages and
limitations. We have combined a wide-range HiRAS spectrum with
higher-resolution spectra recorded at the Learmonth and San Vito
USAF RSTN stations to enhance their quality.
Figure~\ref{F-type_II_spectrum}b presents the combined spectrum
along with higher-sensitivity fixed-frequency records from the
Learmonth RSTN radiometers and \textit{Nobeyama Radio
Polarimeters} (NoRP; \opencite{Torii1979}) at 1 GHz. Their
pre-burst levels correspond to their frequencies, and the peak
fluxes are specified just after the bursts. The bursts correspond
to a faint drifting continuum suggested by the HiRAS spectrogram.
For comparison Figure~\ref{F-type_II_spectrum}a shows the
acceleration of the flux rope and the HXR burst (similar to
Figure~\ref{F-rope_measurements}d). The vertical dashed line
denotes the wave onset time, $t_0 = $~05:35:10.

Dynamic spectra generally present superposition of emissions, which
originate at different sites. The combined spectrum reveals a
complex multi-lane structure of the type II burst that is barely
visible in individual spectrograms. To understand this structure, we
outline the trajectories of separate lanes with the curves of
different line styles and colors. The `F' subscripts denote the
fundamental and `H' the harmonic emission. All of the curves
correspond to a single shock with an onset time $t_0$, crossing
various coronal structures located in different directions from the
wave origin. The different curvatures of the trajectories are most
likely due to different plasma density falloffs, $\delta$, in the
corresponding directions.

The fixed-frequency data reveal the early wave signatures before the
onset of the type II burst. The outlining curves cross them at the
onset or rise of the bursts at fixed frequencies, if the radiometers
were sensitive enough to detect their fluxes, all of which did not
exceed 30~sfu, being $< 10$~sfu below 700~MHz. The type II burst
started at 05:37:00, when the wave front has already passed through
the rim and expanded farther away. The harmonic type II emission was
much stronger than the preceding drifting continuum and reached
191~sfu.

The initial trajectories of the paired type II bands 1 and 3 are
outlined with density falloff exponents of $\delta_1 = \delta_3 =
2.05$, which are close to $\delta = 2.1$ found in
Section~\ref{S-Wave} for the propagation of a spherical wave along
the solar surface. The fundamental emission is visible only
occasionally. Then the bands had turns, which the outlining curves
emphasize. At 05:43:00, bands 1 and 3 passed into bands 2 and 4,
respectively ($\delta_2 = \delta_4 = 2.58$). Another turn occurred
between 05:48:00 and 05:49:30 from harmonic bands $2_\mathrm{H}$ and
$4_\mathrm{H}$ to bands $5_\mathrm{H}$ and $6_\mathrm{H}$,
respectively ($\delta_5 = \delta_6 = 2.75$). The corresponding
fundamental bands left the observed frequency range. The bands
before and after the turns overlapped.

The $\delta = 2.58$ is not much different from $\delta = 2.5$ found
in Section~\ref{S-CME_Formation} for the wave propagation away from
the Sun, being close to the equatorial Saito model
(\opencite{Saito1970}; see also \opencite{Grechnev2011_I}). A nearly
radial direction is also appropriate for $\delta = 2.75$,
corresponding to the Saito model at a higher latitude.

These facts suggest that the type II emission originated initially
at a lower-latitude flank of the wave front, and then, possibly, at
both flanks, while the sources moved away from the Sun. A number of
studies converge to the idea that a probable source of a type II
burst is the current sheet of a small coronal streamer stressed by a
shock front. This causes a flare-like process running along the
streamer (\textit{e.g.}, \opencite{Uralova1994};
\opencite{Reiner2003}; \opencite{MancusoRaymond2004}). A large-scale
shock front crossing a wide range of plasma densities in the corona
can only produce a drifting continuum \cite{KnockCairns2005}; the
narrow-band harmonic emission can originate in a distinct extended
narrow structure like a coronal ray. This scenario accounted for
some structural features of type II bursts observed in different
events (\citeauthor{Grechnev2011_I}, \citeyear{Grechnev2011_I,
Grechnev2011_III, Grechnev2014_II, Grechnev2015}). Imaging
meter-wave observations of type II sources support this scenario
\cite{Feng2013, Chen2014, Du2014}.

The split bands 1--2--5 and 3--4--6, which are better visible in
the harmonic emission (H) in Figure~\ref{F-type_II_spectrum}, have
been interpreted in terms of the emissions upstream and downstream
of the shock front \cite{Smerd1974}. The dynamic spectrum shows an
additional harmonically related equidistant feature
(F$1_\mathrm{F}$, F$1_\mathrm{H}$), which resembles the split
bands in the slope at that time. However, the traditional
interpretation cannot account for a third paired band.
\inlinecite{Du2014} presented observations of type II bursts also
challenging to this interpretation. \citeauthor{Grechnev2011_I}
(\citeyear{Grechnev2011_I, Grechnev2015}) proposed that the
band-split type II bursts could be due to emissions from two
nearby streamers. The complex structure of the type II burst in
this event could be due to both scenarios. The
\inlinecite{Smerd1974} scenario might correspond to the paired
bands 1--2--5 and 3--4--6, and not to the (F$1_\mathrm{F}$,
F$1_\mathrm{H}$) feature. Note that, at least, one streamer is
necessary in either scenario to get a narrow-band emission, which
is crucial to make the band-splitting detectable.

The results of this section quantitatively agree with the wave
development revealed in Sections \ref{S-CME_Formation} and
\ref{S-Wave} and show that some of the considerations and
conclusions of the preceding studies need refinement. Relating the
type II emission source to the region ahead of the CME nose,
invoking the bow-shock properties for the estimations of coronal
parameters on the way of the wave front, the assumption of its
cylindrical geometry, and some others are among them.

\section{Discussion}
 \label{S-Discussion}

The 13 June 2010 event presents a rare case when the kinematical
measurements of the features detectable inside the developing CME
allows one to figure out the formation of its structure. The
initiator of the event was a small eruptive filament. The outcome of
the eruption was the appearance of an expanding coronal wave, whose
front resembled a semi-sphere, with a spheroidal cavity inside
bounded by the rim. From the observational point of view, this event
was strikingly similar to the SOL1996-12-23 event addressed by
\inlinecite{Dere1997}, who demonstrated for the first time the
development of a large-scale CME from a small volume. However, the
active role in our event of the eruptive filament located inside the
cavity was underestimated.

This view was probably the major reason to interpret this event in
terms of a popular concept, which assigns to the coronal cavity a
key role in the coronal wave excitation and creation of the CME
itself. This concept considers the cavity as a cross-section of a
large magnetic flux rope, a major driver of a CME. The filament
(prominence) is regarded as a passive element embedded in the
flux-rope structure. The filament eruption is considered, at most,
as a trigger destabilizing the large flux rope. This view is
popular, although filaments erupting from active regions resemble
small flux ropes, and their behavior does not depend on their
large-scale environment. The filaments expand during the
acceleration stage earlier and sharper than other structures;
their shapes can rapidly change, according to their small sizes.
The impression of the dominant role of a large flux rope, most
likely illusive, probably appeared due to relatively slow
eruptions of extended filaments outside of active regions, as
discussed in \citeauthor{Grechnev2015} (\citeyear{Grechnev2015},
Section~4.1).

The traditional concept of a flux rope rooted in the photosphere
only by two of its ends simplifies the magnetic structure of
observed filaments. A real filament has additional lateral
connections to the photosphere by numerous threads. It is not
clear what is the surface of the magnetic rope related to a real
filament. This surface cannot exceed the magnetic domain
containing the pre-eruption filament. Most likely, the cavity in
the 13 June 2010 event was bounded by the magnetic domain
containing the pre-eruptive filament, because its photospheric
base was fixed, being confined by the separatrix surfaces.

In the traditional view, the only source of a coronal wave is the
outer surface of the large flux rope, which acts as an expanding
piston. The rim in our event looks very similar to this surface
that led to their unfounded identification in previous studies. A
popular assumption was also postulated that the wave can only
become a shock, if the cavity (rim) expands with a super-Alfv{\'
e}nic speed. However, this bow-shock scenario is not the only
possible mechanism of the shock-wave excitation
\cite{VrsnakCliver2008}.

\subsection{What Was the Driver of the CME and Coronal Wave?}
 \label{S-driver_CME_wave}

There are two options: either a large flux rope, which occupies
the whole volume of the cavity, or a small flux rope formed from
the erupting filament inside the cavity. These two options seem to
be very similar. A basic question related to the CME formation is
what drove the wave at different stages of the event. In other
words, we ask if the surface of the piston was inside the cavity
or at its boundary, and if the transition between the two options
was possible.

The primary source of any motions in the event was a small
eruptive filament located deep inside the active region's magnetic
core. The transmission of the motions from the filament outward
was obviously wavelike, as Figures \ref{F-CME_formation},
\ref{F-history_36_deg}, and \ref{F-kinematics_loops} demonstrated.
The wave front appeared conspicuously earlier than the rim, not
\textit{vice versa}, as one might expect from the traditional view
on the cavity. The rim appears to be formed as the approach of the
trajectories of plasma structures 1--4, which started expanding
after the passage of the wave front. The earlier a structure
started to move, the earlier it disappeared because of a rapid
decrease of the emission measure in expanding loops. Eventually,
the rim became the outer envelope of the loops, whose behavior was
similar.

The wave trajectory in Figure~\ref{F-history_36_deg} originates at
the `Rope' trajectory corresponding to the top of the eruptive
filament. Later it decelerated, but the wave trajectory did not
respond to this kinematical change. Thus, the role of the piston
was transferred from the eruptive filament to the forming rim. By
connecting the trajectories of the observed flux rope and the rim
with a single line corresponding to the so-called virtual piston,
we get a solution of a known single-piston problem (see,
\textit{e.g.}, \opencite{Sedov1981}). The trajectories of the wave
front and virtual piston intersect, when the piston was the
filament-associated rope. This confirms that the primary source of
the wave was the observed flux rope developed from the erupting
filament inside the cavity, and not the rim, \textit{i.e.}, its
outer boundary.

In summary, the role of the major piston responsible for the wave
excitation and formation of the rim was played by the small
erupting filament, which occupied the central part of the cavity.
As time elapsed, this piston dilated, acquiring a clinging
magnetic shell, and the whole volume of the cavity became the
virtual piston, whose surface became the rim. This scenario seems
to present consensus between the two different concepts.

We do not consider the wave excitation scenario by the flare
pressure pulse. In a solar flare occurring due to magnetic
reconnection, it does not seem possible to produce the plasma
pressure considerably exceeding the magnetic pressure. For this
reason, the increase in the volume of flare loops is insufficient
to produce an appreciable MHD disturbance outward (see,
\textit{e.g.}, \opencite{Grechnev2015}; for more detail, see
\citeauthor{Grechnev2006beta}, \citeyear{Grechnev2006beta,
Grechnev2011_I, Grechnev2014_II}). The plasma density and
temperature in flare loops are manifested in their SXR emission.
It is intrinsically gradual, resembling the antiderivative of the
HXR burst (the Neupert effect; \opencite{Neupert1968}), which
roughly corresponds to the acceleration of an eruption responsible
for a strong MHD disturbance. As Figures
\ref{F-rope_measurements}c and \ref{F-rope_measurements}d show,
the flare in this event has not yet developed, when the wave
appeared.

\subsection{What Was the Rim?}
 \label{S-rim}

Two observational facts indicate a close association between the
rim and a separatrix surface bounding the magnetic domain, in
which the pre-eruption filament resided. i)~There is a visible
shear between magnetic structures inside the rim and outside it
(Section~\ref{S-CME_lift-off}). ii)~The size of the photospheric
base of the cavity did not change in the course of the eruption.

\begin{figure} 
  \centerline{\includegraphics[width=\textwidth]
   {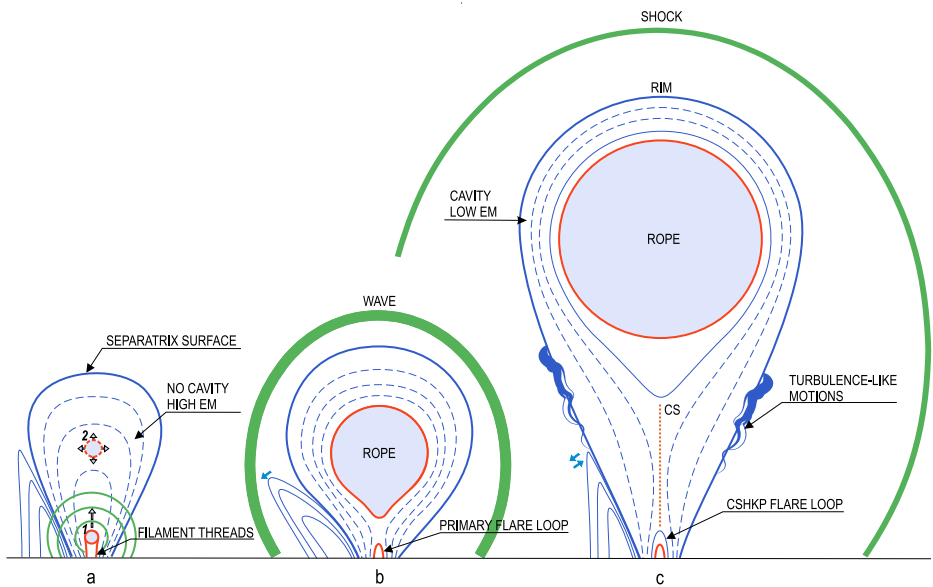}
  }
  \caption{A presumable scenario of the observed event, in which
the major driver of the flare, coronal wave, cavity, rim, and CME
was a small eruptive filament-associated rope. The cartoon
presents a cross section of the active region and the developing
CME at three instances. See text for details.}
  \label{F-scenario}
  \end{figure}

There are additional important indications. iii)~Among the loops
visible in two dimensions (2D) in the \url{AIA_131_171_loops.mpg}
movie, their envelope only coincided with the boundary of the
cavity. This situation reflects the fact that in the 3D geometry,
the outer envelope of the loops belonging to a single domain is
its separatrix surface. iv)~A turbulence-like wave trail is
expected running along the separatrix surface, following the
rising spheroidal cavity. The wave trail should appear due to
plasma motions in 3D loops belonging to adjacent magnetic domains.
These structures should deviate aside and back, as
Figure~\ref{F-scenario} schematically shows. A trail running along
the rim is really visible in Figure~\ref{F-scheme} (circled) and
in the running-difference \url{AIA_131_211_dist.mpg} movie (the
asymmetric arc in the movie connects the top of the eruption with
the trails on both sides). The trail running along the south part
of the rim is indicated by the arrow, and the north trail
resembles turbulence. These features altogether strengthen an
impression of an oblate shape of the rim.

Intriguing was the idea of \inlinecite{Patsourakos2010} that `the
lateral overexpansion may well be the process through which
eruptions starting small in the corona become large-scale CMEs
further out'. However, this observed phenomenon does not seem to
be a physically significant factor for the CME formation for the
following reasons.

i)~As shown in Section~\ref{S-CME_formation_in_EUV}, the decrease
of the aspect ratio in Figure~\ref{F-kinematics_loops}d,
characterizing the apparent ellipticity of the rim, most likely,
was a reversible temporary effect. One of its causes could be a
dampened twisting rotation and writhe of the erupting
filament-related rope inside the cavity. Indeed, its liftoff
direction rapidly turned during the temporary decrease in the
aspect ratio.

ii)~The apparent ellipticity of the cavity in the 13 June 2010
event might be due to its observed geometry with some extent
inclined to the line of sight. Note that the cavity in the analog
of our event presented by \inlinecite{Dere1997} in their Figure~2
was perfectly spherical.

iii)~One might relate the increasing ellipticity of the cavity to
the bow-shock regime, when the dynamic pressure on the frontal
part of the body is important. However, the bow-shock regime is
ruled out by the dissimilar evolutions of the wave and rim (CME)
speeds in Figure~\ref{F-history_22_deg}b.

iv)~As \inlinecite{Grechnev2011_III} showed, the shock
front can be oblate in the radial direction due to the fast-mode
speed distribution. This effect can also contribute to the
impression of the lateral overexpansion (see, \textit{e.g.},
Figure~\ref{F-aia211_wave}b).

Thus, a number of different reasons can result in a visual
effect of the lateral overexpansion. In any case, the varying
shape of the arcade loops and the rim, which were initially
passive, presents their response to the disturbances produced by
the flux rope inside them. These variations are of a secondary
importance, in contrast to the hypothesis relating the outer
surface of the active flux rope to the rim.

\subsection{Overall Scheme}

Figure~\ref{F-scenario} summarizes the results of our analysis of
the 13 June 2010 event, starting from the onset of the eruption.
The initiation phase with a long-lasting filament heating is not
considered. A new key item in this scheme is a dominant, rather
than passive, role of a small filament in the formation of the
classical structural CME components. These are the coronal cavity,
the rim with the frontal structure, and the coronal wave. The
filament associated with a core of this configuration determines
the subsequent evolution of all these components.

Figure~\ref{F-scenario} presents a vertical cross-section of the
active region's magnetic core. The base corresponds to the
photosphere. The thin solid and dashed lines represent the
magnetic field lines. The thick blue line is a separatrix bounding
the major magnetic domain. Figure~\ref{F-scenario}a presents in
the lower part of the domain a pre-eruption filament~1 (small red
circle), which is a rope-like structure rooted in the photosphere
by its lateral threads (red). There is no visible cavity in the
initial state; the plasma emission measure in the major domain is
rather high. The electric current flowing along the separatrix
surface is insignificant. Magnetic loops of an adjacent domain are
shown on the left.

Reconnection between the filament threads creates a primary flare
loop (red in Figures \ref{F-scenario}b and \ref{F-scenario}c) and
primary ribbons, heats the filament, and transforms it into a flux
rope (contoured by the red line), which is not connected to the
photosphere laterally. Then the flare develops as in the CSHKP
model.

Figures \ref{F-scenario}a and \ref{F-scenario}b show the expansion
and lift-off of the flux rope in two stages. (a)~The rope rises,
keeping its size and a large internal magnetic pressure. (b)~Then
the rope expands, equalizing the internal and external pressure.
The lift-off process consists of many such steps. The deformations
corresponding to each step are transferred outward by fast MHD
waves (green ovals in Figure~\ref{F-scenario}a). Each following
wave overtakes a preceding one. The front of a resulting
disturbance can become a shock discontinuity (green oval in
Figure~\ref{F-scenario}c).

The blue dashed lines between the red circle and thick blue
separatrix represent the loops. The expansion and stretch of the
loops surrounding the rope is accompanied by their pressing to
each other and a decrease in their emission measure. The loops
sequentially disappear. The major domain transforms into a cavity
bounded by the separatrix surface observed as the rim. Its
vicinity is, in fact, a moving current sheet separating magnetic
field lines of different connectivity. The current sheet can heat
the surrounding plasma. The blue arrows in the left parts of
Figures \ref{F-scenario}b and \ref{F-scenario}c indicate
deviations of the loops during the passage of the spheroidal
cavity that results in the propagation of a turbulence-like trail
along the separatrix.

This scenario and the cartoon in Figure~\ref{F-scenario}
present the major aspects of the CME and wave development only,
and do not account for all phenomena observed in the event. For
example, the changing lift-off direction and rotation of the flux
rope are not shown. The behavior of the whole system in this case
can be understood, keeping in mind that adjacent magnetic surfaces
slip along each other.

A subsequent story of CMEs and related waves seems to be more or
less clear. Our view, expectations, and results can be found,
\textit{e.g.}, in \citeauthor{Grechnev2014_II}
(\citeyear{Grechnev2011_I, Grechnev2013_20061213, Grechnev2014_II,
Grechnev2015}). The CME should enter the stage of a free
expansion, which is known to be close to self-similar,
\textit{i.e.}, the distances between all of its structural
components progressively increase. Most likely, the free expansion
regime of the 13 June 2010 CME has not yet established within the
AIA field of view. Later on, the loops, which disappeared,
apparently merging to the rim, should reappear and diverge in the
self-similar regime. The expectations are consistent with the
LASCO observations of the CME discussed in Section~\ref{S-lasco}.
For an independent verification of our conclusions related to the
waves, see, \textit{e.g.}, the studies by \inlinecite{Kwon2013};
\inlinecite{Kwon2014}; \inlinecite{Kwon2015}, and others mentioned
in the text.

\subsection{Comparison with Different Observations}

Most of the items of the outlined scenario has been observed in
different events. The long-lasting heating of pre-eruptive
filaments manifesting in the rise of the SXR flux, firstly stated
by \inlinecite{Zhang2001}, seems to be rather common. It was shown
explicitly or implicitly by, \textit{e.g.},
\inlinecite{Kundu2009}, \inlinecite{Meshalkina2009}, and
\citeauthor{Grechnev2014_I} (\citeyear{Grechnev2014_I,
Grechnev2015}). Erupting filaments often become bright in 171 or
195~\AA\ which indicates their heating up to, at least, 1~MK (see,
\textit{e.g.}, the movies at
\url{http://trace.lmsal.com/POD/TRACEpod.html#movielist}). Their
brightness in 195~\AA\ well before the flare peak occasionally
exceed the flare arcades observed later \cite{Chertok2015}.

A completely formed flux rope identical to the erupted one
unlikely pre-exist because of a large excess of the reconnected
over pre-existing poloidal flux \cite{Qiu2007, Miklenic2009} and a
strong related propelling Lorentz force. Genesis of eruptive flux
ropes from hot parts of filaments or, at least, their association
has been reported \cite{Kumar2012, Cheng2013,
ChenBastianGary2014}. A flux-rope progenitor is not always
observed as a filament, probably, due to its higher temperature
and lower density, but this should not affect the flux-rope
formation.

Erupting filaments and associated hot flux ropes vigorously twist,
writhe, and expand sharper and earlier than any surrounding
structures. This demonstrates their role as internal drivers of
the CME formation process (\citeauthor{Cheng2011}
\citeyear{Cheng2011}, \citeyear{Cheng2013}; \opencite{Zhang2012}).
The deformations in the magnetic configuration caused by the
erupting flux rope are transferred to the environment by a
fast-mode MHD wavelike disturbance propagating outward, whose
manifestations can be detected far outside. A static coronal
structure pushed by the wave unlikely can drive this wave just
afterwards. The change to the piston-driven shock is possible
later, when the regime of plasma extrusion by the CME bubble
changes to the plasma flow around it, if the CME speed exceeds the
ambient fast-mode speed \cite{Grechnev2015}.

To verify our results and check if they really help in
understanding eruptive phenomena, we briefly consider the
SOL2010-11-03 event. Its flare site at S19\,E98 was partly
occulted. This largely reduced the flare emission with an
estimated importance of M5.8 \cite{Chertok2015} to the actually
observed C4.9 level, favoring the observations of an erupting flux
rope, but hid its early development. \inlinecite{Cheng2011}
revealed a hot flux rope, which rapidly moved outward and
compressed the surrounding coronal plasma to form the pileup
observed as a bright rim in lower-temperature channels, similar to
our event.

\inlinecite{Zimovets2012} measured the average speed of the flux
rope within the AIA field of view of $\approx 500$~km~s$^{-1}$ and
the leading edge of the EUV transient of up to 1400~km~s$^{-1}$.
On the other hand, the white-light CME was as slow as
241~km~s$^{-1}$ according to the CME catalog. This fact not
considered previously shows that the CME decelerated even more
than in our event.

Nevertheless, the event produced a shock wave, which manifested in
a band-split type II burst. From the position of its source
observed in radio images above the CME top, \inlinecite{Bain2012}
proposed that the shock with an estimated speed of
1900--2000~km~s$^{-1}$ was piston-driven (that is difficult to
reconcile with a low speed of the white-light CME). Conversely,
\inlinecite{Zimovets2012} concluded that the shock wave was
initially driven by the eruptive plasmas, but later transformed to
a freely propagating blast shock wave, that is very close to the
scenario we describe. They also concluded that the band splitting
was preferably due to emissions from the regions
upstream/downstream of the shock.

\inlinecite{Bain2012} and \inlinecite{Zimovets2012} did not
consider a cause of the narrow bandwidth of the type II burst,
especially important to reveal the split bands. As shown in
Section~\ref{S-type_II}, this is only possible if its source is
compact, being located, \textit{e.g.}, in a coronal streamer.
Unlike our event, where the probable type II source was a remote
streamer hit by a quasi-perpendicular shock, the type II emission
in the SOL2010-11-03 event seems to have been produced by a
quasi-parallel shock in the main streamer above the parent active
region. This situation was discussed by
\citeauthor{Grechnev2014_II} (\citeyear{Grechnev2014_II};
Figure~10); its particularity is a considerably higher drift rate,
which was really the case in the SOL2010-11-03 event.

\section{Conclusions}
 \label{S-Conclusion}

Taking advantage of the detailed multi-instrument observations of
the 13 June 2010 event with an unprecedented temporal and spectral
coverage, primarily thanks to the SDO/AIA data, it has become
possible to reveal a consistent picture of a solar eruption,
coordinating qualitatively and quantitatively its various
observational aspects. The inferred scenario updates and specifies
existing hypotheses. Unlike traditional expectations, the major
driver of the flare, CME formation, and a large-scale wave was the
erupting filament. It heated up to 10~MK and even higher; being a
direct progenitor of a hypothesized flux rope, the filament
transformed into a bundle of erupting loops, which sharply
expanded and thus produced a strong MHD disturbance inside the
developing CME. This outward-propagating disturbance passed
through the forming CME and ran ahead of all its structures.
Probably, the disturbance rapidly steepened into a shock
resembling a blast wave, produced a type II burst and EUV wave.
The magnetic domain containing the eruptive filament/flux rope was
forced to expand from inside and became the CME cavity bounded by
a separatrix surface observed as a rim. With a nearly spheroidal
or pyriform shape, its appearance depends little on the flux-rope
orientation. The enhanced-temperature coronal loops above were
swept up by the expanding rim and became the CME frontal
structure.

Although the envisioned identity of the cavity with a flux-rope
has not been confirmed, the role of its rim was important. Being
not permeable for plasma, the expanding separatrix surface
associated with the rim was responsible for the appearance of
dimming in the rarified volume behind it, and took the role of a
piston after the deceleration of the flux rope. If the CME were
fast, then the wave ahead it eventually changed to the bow-shock
regime and became the CME-driven shock in a correct sense. This
has not happened, because the CME was slow; the decelerating wave
dampened and decayed into a weak disturbance.

In spite of a similarity between the extremities of the bow-shock
and blast-wave regimes, some of their properties are different. In
particular, when the blast wave is getting weaker, then the
distance between its front and the piston decreases. The relation
is opposite in the bow-shock regime. For this reason, assumptions
of the bow-shock properties for the waves impulsively excited by
eruptions might result in an incorrect outcome.

The preceding studies of the 13 June 2010 eruptive event have
revealed its important aspects. However, the flux rope, which was
a key item of the eruptive process, escaped detection. Due to this
difficulty, the researchers had to invoke some traditional
assumptions, not all of which have been confirmed. It was also
very difficult to detect the flux rope for us. It has become
possible, because our persistent search was guided by the
expectations based on our preceding results (\textit{e.g.},
\citeauthor{Grechnev2011_I}, \citeyear{Grechnev2011_I,
Grechnev2014_I, Grechnev2014_II, Grechnev2015}), and in this way
we elaborated the data processing and analysis techniques outlined
in Section~\ref{S-Methodical_Issues}. These techniques might also
be helpful in different studies.

Most likely, the updated scenario of an eruptive event presented
here is rather typical. Incorporating its items into theoretical
considerations and numerical simulations seems to be promising for
a better understanding solar eruptive phenomena. Observational and
theoretical studies of the causes and mechanisms of the
long-lasting pre-eruptive heating, which we did not analyze, could
help in perceiving the triggers of solar eruptions and their
practical forecasting.

\begin{acks}
We thank S.A.~Anfinogentov, N.S.~Meshalkina, and M.V.~Eselevich
for their assistance in data handling; and A.~Kouloumvakos with
co-authors and M.A.~Livshits for useful discussions. We are
indebted to an anonymous reviewer for the cooperation in bringing
this article to its final form. We thank the instrumental teams
operating AIA and HMI on SDO, STEREO/SECCHI, \textit{Fermi},
SOHO/LASCO, Nobeyama Solar Facilities, NICT, USAF RSTN network,
and GOES satellites for the data used here. The CME catalog used
in this article is generated and maintained at the CDAW Data
Center by NASA and The Catholic University of America in
cooperation with the Naval Research Laboratory. SOHO is a project
of international cooperation between ESA and NASA. This study was
supported by the Program of Basic Research No. II.16.1.6, the
Integration Project of RAS SD No.~4, and the Russian Foundation of
Basic Research under grants 15-02-01077, 15-02-01089, and
15-32-20504-mol-a-ved.

\end{acks}

\end{article}

\end{document}